\documentclass[appendixfloats,iop]{emulateapj}
\usepackage{longtable}
\usepackage{xcolor}
\begin{document}

\title{A Catalog of the Most Optically Luminous Galaxies at $z<0.3$: \\
        Super Spirals,  Super Lenticulars,  Super Post-Mergers, and Giant Ellipticals}

\author{Patrick M. Ogle$^1$, Lauranne Lanz$^2$,  Philip N. Appleton$^3$, George Helou$^3$, Joseph Mazzarella$^3$}

\affil{$^1$ Space Telescope Science Institute, 3700 San Martin Drive, Baltimore, MD 21218}

\affil{$^2$ Dartmouth College, NH}

\affil{$^3$IPAC, California Institute of Technology, 
       Mail Code 220-6, Pasadena, CA 91125}

\email{pogle@stsci.edu}

\shorttitle{Super Spirals}
\shortauthors{Ogle et al.}

\begin{abstract}

We present a catalog of the 1525 most optically luminous galaxies from the Sloan Digital Sky Survey
(SDSS) with $r$-band luminosity $L_r>8L^*$ and  redshift $z<0.3$, including 84 super spirals, 15 super lenticulars, 14 super post-merger galaxies, and 1400 giant ellipticals.
With mass in stars of $10^{11.3}-10^{12} M_\odot$, super spirals and lenticulars are the most massive disk galaxies currently known.  The specific star formation rates 
of super spirals place them on or below the star-forming main sequence. They must have formed stars at a high rate throughout their history in order to grow their 
massive, gigantic stellar disks and maintain their blue $u-r$ integrated colors.  Their disks are red on the inside and blue on the outside, consistent with inside-out growth. They 
tend to have small bulge-to-total ($B/T$) $r$-band luminosity ratios, characteristic of disk building via minor mergers and cold accretion.  A large percentage of super disk galaxies 
(41\%) have double nuclei, double disks, or other signatures of ongoing mergers. Most (72\%) are found in moderate to low density environments, while the rest are
found at the outskirts of clusters.  It is likely that super spirals survive in these environments because they continue to accrete cold gas and experience only minor mergers at late times, 
by virtue of their enormous masses and angular momenta. We suggest that  super post-mergers are the product of super-spiral major mergers and
may be the precursors of some giant elliptical galaxies found in low density environments.  We present two new gravitational lens candidates in the Appendix.

\end{abstract}

\section{Introduction}

We recently found that $\sim6\%$ of the most optically luminous galaxies at redshift $z<0.3$ are giant, high-surface brightness spiral galaxies, with masses of $10^{11}-10^{12} M_\odot$
and isophotal diameters of 55-140 kpc \citep{oln16}.  These super spiral galaxies are actively forming stars and appear to be vastly scaled-up versions of normal spiral galaxies.   
The extreme sizes, masses, and luminosities of super spirals extend the parameter space over which galaxy scaling laws may be studied, providing a new arena to test theories
of massive galaxy formation and evolution. 

In addition to giving new insights into galaxy formation and growth, super spirals can help discriminate among proposed mechanisms for quenching star formation in massive galaxies.  
The red optical colors of massive spiral galaxies may indicate that they are starved of gas and dying \citep{bnb09,mmr10,sus14}.  However, it is important to determine for any individual galaxy
whether its red colors indicate low specific star formation rate or dust extinction via mid-infrared (MIR) photometry \citep{fbp16}.  Star formation is thought to be quenched in most galaxies above a 
mass in stars of $\sim 10^{11} M_\odot$, by collisions, AGN activity, accretion shocks, or ram-pressure stripping, which turn them into red and dead elliptical or lenticular galaxies 
\citep{hhc06,db06,edl07,mbt09,sus14,cvc15}. Though they are rare, the existence of super spirals demonstrates that the limit to spiral galaxy size and mass is much higher than previously thought, 
and that high mass in stars can not be the primary cause of star-formation quenching. In fact, spiral galaxies with mass in stars $\sim 10^{11} M_\odot$ may be the most efficient at converting gas into stars, with mass 
fractions in stars approaching the cosmological baryon fraction \citep{pfm19}.  Super spirals and giant ellipticals may represent distinct evolutionary pathways for the most massive galaxies, depending on 
dark halo mass and angular momentum.  Super spirals may remain unquenched because they reside in less massive dark halos than giant ellipticals of similar mass in stars.

Hydrodynamical simulations that include increasingly realistic star formation and active galactic nucleus (AGN) feedback prescriptions \citep{sh05,g07,hcy09} show that galaxy mergers play a key role in 
determining the configuration, dynamics, star-formation histories, and ultimate fates of galaxies.  Simulations also reveal that typically half of the mass in stars  is acquired via mergers, 
while the rest is formed in situ \citep{rg16}.  A relatively large fraction of super spirals have two bulges surrounded by a common spiral disk or tidal features \citep{oln16}, demonstrating the importance of mergers for 
these most massive disk galaxies.   However, super spirals may survive most mergers because of their extremely large masses and sizes. That is, for a super spiral, even a merger with a typical 
$L^*$ galaxy is a  high mass ratio, minor merger that will not destroy its massive disk. 

This work extends the search for the most massive, optically luminous galaxies, relaxing the restriction that they be detected in the near-UV ({\it NUV}-band) by the {\it Galaxy Evolution Explorer (GALEX)}, 
with the goal of understanding how super spirals relate to other types of massive galaxies.  We present a new class of super lenticular galaxies that may be quenched super spirals and a new
class of super post-merger galaxies that may be the product of super-spiral major mergers.

\section{Sample and Photometry}

\subsection{Sample Selection}
Our galaxy sample is primarily selected for high SDSS $r$-band luminosity ($L_r>8L^*$), which is an excellent tracer of mass in stars for unobscured galaxies. Unlike \cite{oln16}, here we do not  impose 
requirements of {\it GALEX} $NUV$-band detection or spiral morphology, resulting in a sample that also includes lenticular, elliptical, and peculiar galaxies, and which is less biased towards high star formation rates.
We selected all galaxies from the NASA/IPAC Extragalactic Database (NED) with an existing spectroscopic redshift $z<0.3$ and $r$-band photometry from SDSS I or II (\citep{y00,s02}.  These galaxies were then ranked by their $r$-band monochromatic luminosities ($L_r$), after  correcting for Galactic extinction and applying a K-correction.  We report $L_r$ relative to the characteristic luminosity at the knee of the galaxy luminosity function of $L^*=5.41 \times 10^{43}$ erg s$^{-1}$ at 6200 \AA~  \citep{bhb03}.  

The redshift distribution for spiral galaxies with $L_r>8L^*$ cuts off  at $z=0.30\pm0.02$ (Fig. 1), corresponding to the SDSS I/II spectroscopic selection limit of $r=17.77$ \citep{s02}.  The distribution of elliptical galaxies cuts off at a higher redshift of  $z=0.38\pm 0.01$, corresponding to the effective redshift cutoff of the SDSS luminous red galaxy (LRG) sample \citep{eag01}.  Therefore, to mitigate against incompleteness and selection effects at greater redshifts, we restrict the present study to SDSS galaxies with $L_r>8L^*$  at $z<0.3$.

We find 1616 candidate galaxies at $z<0.3$ with $L_r>8L^*$, presented as the Ogle et al. Galaxy Catalog (OGC: Tables A1-A4).   We visually inspected the SDSS 3-color images in order to determine their morphologies and checked their redshifts against their SDSS spectra.  Galaxies were classified as spiral, lenticular, elliptical, or peculiar, based on visual appearance. In particular, relative prominence of bulge and disk components and presence of spiral arms were 
the key discriminants. We give a breakdown of OGC galaxy types in Table 1.  A total of 1525 galaxies are legitimate high-luminosity galaxies.   The remaining 91 galaxies that have inaccurate photometry
(51 in Table A2), or incorrect redshifts (24 in Table A3), or that overlap with foreground objects (16 in Table A4) are excluded from further analysis.

Apart from constructing a manageable sample of the most optically luminous galaxies, there is no particular physical significance to our $8L^*$ cutoff.  As we
shall demonstrate, super spirals are dramatically scaled-up versions of much more abundant $L^*$ spirals, albeit with some significant structural, photometric, and other differences.
Since our $8 L^*$ limit is somewhat arbitrary,  the search for superluminous, giant spiral galaxies could be extended down to lower luminosities. 
For example, there are 16,301 SDSS galaxies with $L_r> 5 L^*$ at $z<0.25$.  Using this $5 L^*$ cutoff would
yield galaxies in the top $2\%$ of the $r$-band luminosity distribution,  compared to the top $0.2\%$ for our adopted $8 L^*$ cutoff.

\begin{deluxetable}{llrr}
\tablecaption{OGC Galaxy Type Percentages}
\tablehead{
\colhead{Type} & \colhead{Subtype} & \colhead{Number} &\colhead{Percentage}}
\startdata
Super Disk &                                     & 99 & 6.5\\
\hline
                      & Spiral                     & 84  & 5.5\\
                      & Lenticular (S0/Sa)  & 15  & 1.0\\
\hline
Giant Elliptical &                               & 1400 & 91.8\\
Post-merger    &                               & 14 & 0.9 \\
Bright AGN      &                                & 12 & 0.8\\
\hline
Luminous galaxies &                                & 1525 & 100.0\\
Rejects              &                                & 91  & \\

\hline
All  &                                 &1616 & 
\enddata
\end{deluxetable}

\subsection{Super Spirals}
 
We find 84 super spiral galaxies with $L_r>8L^*$  at $z<0.3$ (Table A5, Figs. A1-2). This includes 32 new super spirals, augmenting our original sample of 53  \citep{oln16}.  Inspection of an archival HST image (Appendix) led us to remove OGC 0302, reducing the sample size from 85 to 84. 

In order to compare morphologies (Fig. 2), we cross-matched the OGC with the Galaxy Zoo DR1 catalog \citep{lsb11}. We find that only 22/84 super spirals (26\%)  are classified as spirals with $P_\mathrm{spiral}>0.8$ by Galaxy Zoo, 10 are classified as ellipticals (12\%), 50 are classified as uncertain (60\%),  and 2 (OGC 0574 and 586 ) are not classified (2\%).   Super spirals with uncertain Galaxy Zoo classification include  galaxies that appear to have normal spiral morphologies, such as OGC 0065 and 0713 ($P_\mathrm{spiral}=0.57 $ and 0.60), and spirals that are disturbed by mergers, such as OGC 0789  and 1304 ($P_\mathrm{spiral}=0.61$ and 0.29).  It appears that the Galaxy Zoo classifications of many super spirals are rendered uncertain by their relatively high redshifts \citep{bnb09}  and high merger fraction.

The mean number density of super spirals in our sample is  58 Gpc$^{-3}$ within a comoving volume of 7.14 Gpc$^3$, corrected for the  20.3\% sky coverage of SDSS II.  This is 5.5\% of the total number density of high-luminosity galaxies in our sample (1050 Gpc$^{-3}$).  Correcting for an inclination incompleteness of 40\% (Section 5), the number density of super spirals increases to  97 Gpc$^{-3}$, 9.2\% of the population of galaxies with $L_r > 8 L^*$.  
We emphasize that super spiral galaxies are one of the rarest galaxy populations in the universe. Their comoving number density is a factor of $10^{3}-10^{4}$ lower than samples of massive galaxies constructed from smaller surveys at higher redshift \citep[e.g.,][]{dhm17,fcc17}.  Existing deep surveys are not large enough to yield significant numbers of super spirals.  Deeper large surveys by {\it Euclid}, the {\it Wide Field Infrared Survey Telescope} ({\it WFIRST}), and Large Synoptic Survey Telesope (LSST) will be necessary to discover and characterize super spiral progenitors at $z>1$.

Extending our search for high-luminosity SDSS galaxies to $0.3<z<0.6$ yields 83 additional super spirals (included in Figure 1 but not tabulated).
The redshift distribution of super spirals falls off at $z>0.3$, corresponding to the SDSS I/II magnitude limit for spectroscopy and our $L_r>8L^*$ luminosity lower limit.

\subsection{Super Lenticulars and Giant Ellipticals}

 \cite{oln16} predicted a new class of super lenticular galaxies, but for the most part they were excluded from their sample due to an {\it NUV}-band selection criterion. Here we identify 15 super lenticular (S0/Sa)
galaxies with  $L_r>8L^*$  at $z<0.3$  (Table A5, Figs. A1-2).  It is easiest to identify lenticulars at intermediate inclination (Section 5), where they are less likely to be confused with ellipticals or edge-on spirals.   Based on the small number in our sample, the space density of super lenticulars is $>10$ Gpc$^{-3}$.  It is likely that we are missing $>60\%$ of super lenticulars, at both low and high inclination, based on the distribution of observed inclinations (Section 5).

There are 1400  giant elliptical galaxies in the OGC that  constitute 91.8\% of the sample. Their mean comoving number density is 970 Gpc$^{-3}$, corrected for SDSS II sky coverage.  The most optically luminous giant elliptical galaxy is OGC 0021 (2MASX J12220526+4518109), with $L_r=19.8 L^*$.  It resides at the center of a rich galaxy cluster, which is frequently (but not always) the case for the giant ellipticals in our sample.

\subsection{Super Post-Mergers and Luminous AGN Hosts}

There are 14 peculiar galaxies (0.9\% of the sample) that have disturbed morphologies, indicative of recent mergers (Table A5 \& Fig. A3). We highlight one additional peculiar starburst galaxy (OGC 1662, $L_r=7.9 L^*$)
that does not quite make our $8L^*$ cutoff but has a very peculiar morphology and multi-hued appearance. We include all of these galaxies in our photometric analysis, but differentiate them from galaxies with more regular morphologies.

We exclude 12  non-spiral galaxies (0.8\% of the sample) that are quasar hosts, blazar hosts, or contain bright stars seen in projection from further analysis because an AGN or foreground star is likely to contribute a significant fraction of the luminosity at wavelengths of interest (Table 6 \& Fig. A3).

\begin{figure}[h]
   \includegraphics[trim=2.0cm 7.7cm 3.0cm 9.0cm, clip, width=\linewidth]{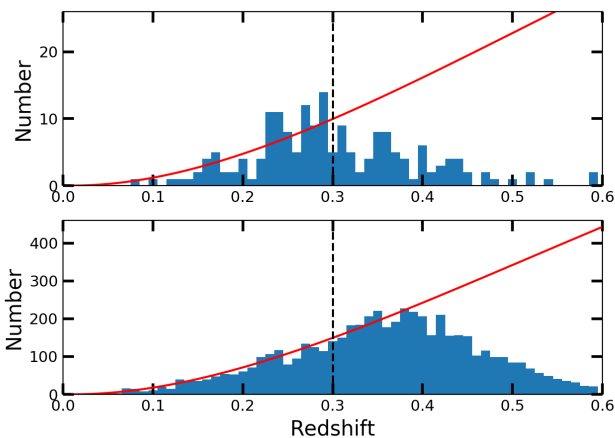}
   \figcaption{Redshift distributions for galaxies with $L_r>8 L^*$.  Top: super disk (spiral and lenticular) redshift histogram compared to a constant comoving density curve. 
   		    Bottom: giant elliptical redshift histogram. The OGC is limited  to $z<0.3$ (dashed line). The distribution for ellipticals extends to higher redshift because they are augmented by the SDSS LRG sample.                  
    \label{1}}
\end{figure}

\begin{figure}[h]
   \includegraphics[trim=2.5cm 3.0cm 3.0cm 3.2cm, clip, width=\linewidth]{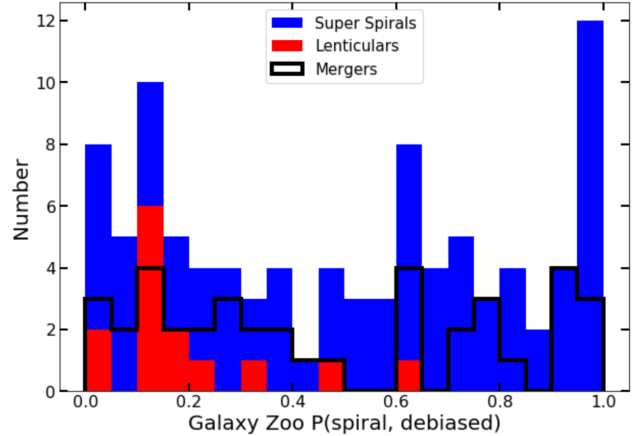}   
   \figcaption{ Galaxy Zoo 1 debiased spiral probability \citep{bnb09}, which is corrected for redshift bias. Because the redshift bias corrections are large for galaxies at $z>0.08$, and spiral galaxies at these redshifts may be misclassified 
    as  early types, Galaxy Zoo morphologies should be interpreted with caution. Total bar height indicates  total spirals and lenticulars, while color  represents the fraction of each type. Black histogram shows the number of mergers, regardless of morphological type.
    \label{2}}
\end{figure}

\subsection{Photometry}

We use CModel catalog photometry from SDSS DR6 \citep{y00}  and aperture photometry from the 2-Micron All-Sky Survey \citep[2MASS,][]{s06} and the {\it Wide-field Infrared Survey Explorer} \citep[{\it WISE},][]{w10} to estimate the total optical luminosities, mass in stars, and star formation rates (SFRs). The smallest 2MASS and {\it WISE} apertures that encompass the $D_{25}$ isophotal diameter at $r=25$ mag were selected. The SDSS CModel magnitudes are  effectively aperture-matched  and are therefore appropriate for measuring integrated galaxy color. All galaxies are detected in the SDSS $u$, $g$, $r$, $i$, $z$ bands.  A total of 90 super disk galaxies  are 
detected in 2MASS $K_\mathrm{s}$ band and 86 in $W3$ band.  Of the 1400 giant ellipticals, 127 are undetected in $K_\mathrm{s}$ band, and 128 are undetected in $W3$.

We correct SDSS and 2MASS photometry for Galactic extinction using the NED extinction calculator, based on the Galactic extinction maps of \cite{sf11}. We K-correct SDSS and 2MASS magnitudes to rest frame values with a simple, custom procedure that performs log-linear interpolation of the  observed spectral energy distribution (SED).  We K-correct {\it WISE} magnitudes using two representative model SEDs, one for star-forming galaxies with $W2-W3 >2$, and a second for quiescent galaxies with $W2-W3 \leq 2$, yielding K-corrections of $K(W3)<0.33$ mag and $K(W2-W3)<0.21$ over the redshift range of our sample. 

\section{Mass in Stars and Star Formation Rate}

The SEDs of nearly all galaxies in our sample are dominated by  stellar populations at NUV to NIR wavelengths, and warm dust emission at mid-IR wavelengths. Galaxies with SEDs dominated by AGNs  (0.8\% of
the sample) are excluded from this analysis. We estimate mass  in stars (Fig. 3) from K-corrected 2MASS $K_\mathrm{s}$ magnitude and K-corrected SDSS $u-r$ color, using the prescription of \cite{bmk03}, with a small correction ($+0.004$ dex) to convert to a \cite{c03} initial stellar mass function (IMF).  The mass estimates are relatively insensitive to both dust extinction and variations in mass-to-light ratio $M/L$ with mean stellar population age.  

\begin{figure}[h]
   \includegraphics[trim=2.0cm 0.0cm 1.0cm 0.0cm, clip, width=\linewidth]{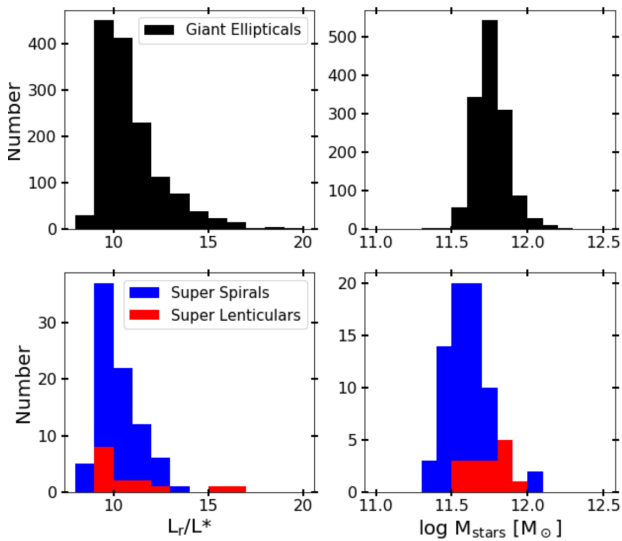} 
   \figcaption{ Luminosity and mass of giant ellipticals compared to super disks. Left: $r$-band luminosity distributions. Right: Distributions of mass in stars.  
   \label{3}}
\end{figure}

\begin{figure*}[h]
   \includegraphics[trim=0.0cm 0.0cm 0.0cm 0.0cm, clip, width=\linewidth]{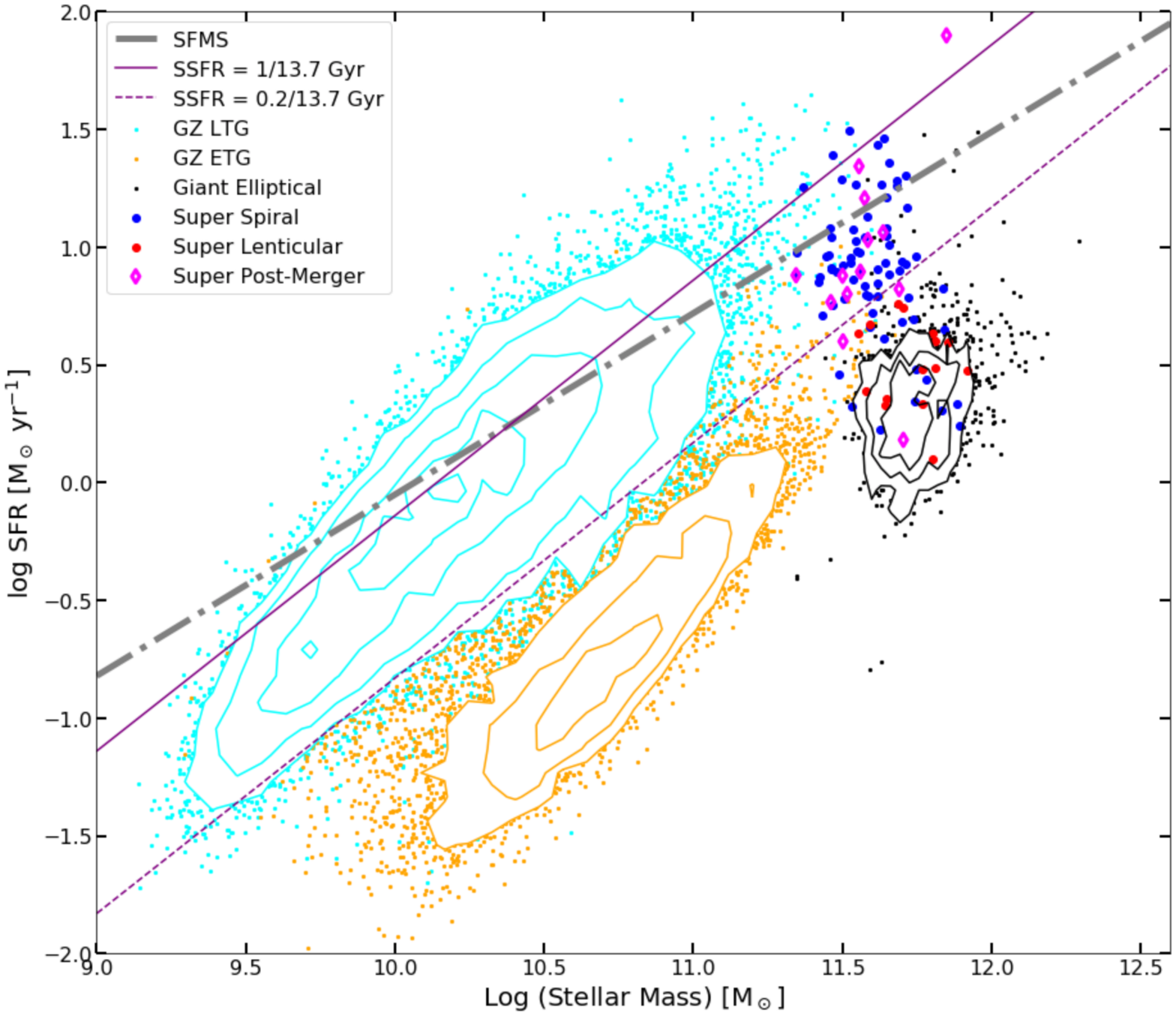}
   \figcaption{
       Star formation rates and masses in stars of super disks, post-mergers,  and giant ellipticals compared to Galaxy Zoo late type and early type galaxies
        \citep[GZ LTGs and ETGs:][]{sus14,asa14}. The thick dot-dashed line indicates  the star-forming main sequence at $z \sim 0$ \citep{edl07}, which has been shifted downward
         by 0.22 dex to match the {\it WISE}-based SFR estimates of  \cite{cvc15}.  The dashed and solid purple lines indicate SSFRs of $(0.2-1)/(13.7$ Gyr), a range where
         active star forming galaxies would double their masses in stars in 1-5 times the current age of the universe.
          \label{4} }
\end{figure*}

The super spiral masses are on average 4 times larger than those given by \cite{oln16}, primarily because of a systematic error they made converting from $K_\mathrm{s}$-band monochromatic luminosity to solar luminosity units.  The $K_\mathrm{s}$ monochromatic luminosity was incorrectly divided by the solar bolometric luminosity, rather than the solar $K_\mathrm{s}$-band monochromatic luminosity, which is a factor of 5.5 smaller. This is partially offset by the new K-corrections to the $K_\mathrm{s}$ band magnitudes, which also act to reduce the scatter in the distribution of masses in stars. These new, more accurate estimates lead to a better understanding of how truly massive super spirals are. They in turn affect the specific star formation rates and location of super spirals relative to the star-forming main sequence (SFMS, Fig. 4), shifting them to the right by 0.6 dex and placing most of them on or below this relation. 

We compare the masses in stars of super disk galaxies and  giant ellipticals in Figure 3.  The narrow range of high $r$-band luminosities in the OGC effectively selects for the most massive galaxies, with masses in stars of $10^{11.3}-10^{12.3} M_\odot$.  The mean mass in stars is a factor of 1.4 (0.14 dex) smaller for super spirals than for  super lenticulars and giant elliptical galaxies, because of their systematically lower stellar $M/L$ ratios.   A major merger between a super spiral and another massive galaxy could potentially create super lenticular and giant elliptical galaxies of greater mass.  Super spiral major mergers may provide a channel to create giant elliptical galaxies outside of galaxy clusters, like the most massive isolated elliptical OGC 0078 (2MASX J02295551+0104361), with $\log M_\mathrm{stars} = 11.9$.  However, the dark halo masses of isolated giant ellipticals like this one are likely to be considerably smaller than the dark halo masses of giant ellipticals in clusters, even though they have comparable masses in stars . The very different evolutionary histories of ellipticals in the most massive halos versus those that form from super spiral mergers may potentially result in different fractional mass in stars, gas content, kinematics, and morphology.  

We estimate SFRs from K-corrected {\it WISE} $W3$-band (12 $\mu$m) fluxes, which trace warm dust heated by UV photons in star-forming galaxies \citep{cvc15}.\footnote{Note that there is a known offset of -0.22 dex  \citep{cvc15} between SFRs derived from {\it WISE} measurements compared to those derived from H$\alpha$ fluxes \citep{edl07, bcw04}.  We adopt the SFMS of \cite{edl07}, shifted downward by -0.22 dex to match {\it WISE}-estimated SFRs.} We plot SFR against mass in stars for OGC galaxies in Figure 4.  This method gives consistent SFR values for star-forming galaxies compared to full-SED fitting with MAGPHYS \citep{dce08, oln16} and is relatively insensitive to dust extinction, compared to SFRs estimated from UV or H$\alpha$ emission.  On the other hand, old stellar populations contribute most of the $W3$-band luminosity for quiescent elliptical galaxies with the lowest SSFRs \citep{cvc15, bpb06}, so their SFRs should be considered to be upper limits. Between these two extremes, both old and young stellar populations will contribute to the $W3$ luminosity.  Assuming a baseline WISE color of $W2-W3=0.32$ for quiescent ellipticals, the break-even point (equal $W3$ luminosity from star-formation and old stars) occurs at  $W2-W3 \simeq 0.32 + 2.5 \log 2 =1.07 $ (Fig. 5).  Only 7 super spirals, 3 super lenticulars, and 1 super post-merger have $W3-W2$ colors bluer than this, such that their $W3$ luminosity is dominated by old stellar populations.  Dust heated by A-type stars in recently quenched galaxies may also lead to an overestimate of the SFR \citep{asa14,abl17}.  This is not the case for most super disk galaxies in our sample, though it may be a concern for the peculiar, post-merger galaxies, depending on the fractions of UV photons from A stars and star-forming regions.

The super spirals and super lenticulars have SFRs ranging from 1-30 $M_\odot$ yr$^{-1} $ (Fig. 4).  Their  SSFRs span a broad range from $(0.02- 1.5)\times 10^{-10}$  yr$^{-1}$ (1/SSFR $=6-500$ Gyr), with most falling on or below the SFMS, extrapolated from less-luminous SDSS spiral galaxies.  There appears to be a continuum of massive, super disk galaxies, from active star-forming super spirals  to quiescent super lenticulars, similar to the well-established sequence for $L^*$ galaxies. There is no sharp dividing line between super spirals and super lenticulars in the $M_\mathrm{stars}$-SFR plane.  Because super spirals are so massive, not even the ones with the highest star formation rates  and infrared luminosities are global starbursts.  Based on SDSS  spectra, we do find strong {\it nuclear} starbursts in 5 super spirals (OGC 0217, 0454, 1457, 1464, and 1520) and relatively strong H$\alpha$ from nuclear star formation in 2 (OGC 1312, and 1512).

The giant ellipticals are for the most part quiescent, with SSFR $<1.5\times 10^{-11}$  yr$^{-1}$.  While a small number (14) have greater SSFRs than this,  they  are not starbursts.  Six of these (OGC 0034, 0087, 0123, 0792, 1412, and 1581) have high Balmer line equivalent widths and significant populations of young, blue stars in their SDSS spectra.  The dust and UV emission from the vast majority of  more quiescent giant ellipticals may indicate a low level of ongoing star formation accompanied by young stellar populations.

Most of the super post-mergers  (11/14) have SSFRs that formally put them on the star-forming main sequence.  However, they have SDSS spectra with high H$\delta$ equivalent widths characteristic of dominant A-star populations in post-starburst galaxies.  The $W3$-band fluxes may therefore have a significant contribution from warm dust heated by UV emission from post-starburst stellar populations  \citep{mp13,abl17}.   NUV emission from A-F type stars can linger for $1-3$ Gyr following starburst activity, leading to significant mid-IR emission. The post-mergers display a range of H$\alpha$ equivalent widths from star formation and AGN activity, indicating that they are still  forming stars and feeding their black holes. Only one (OGC 0792) has an SDSS spectrum that indicates an ongoing burst of nuclear star formation. Five have clear AGN signatures in their SDSS spectra, including two (OGC 0247 and 1413)  with high-luminosity AGNs that contribute significantly to their $W3$-band luminosity and red $W1-W3$ colors.  We may be viewing most of these peculiar galaxies during quenching episodes immediately following gas-rich mergers, where the observed AGN activity may eventually clear out much of the remaining gas.   If all of the star-forming super post-mergers in our sample are the result of super spiral major mergers, we estimate a per-galaxy super spiral destruction rate of 0.16 Gyr$^{-1}$, assuming a post-merger settling timescale of 1.0 Gyr \citep{ljc08}. The two super post-mergers  (OGC 0743 and 1141) with SDSS spectra and SEDs characteristic of predominantly old stellar populations may be the product of dry mergers. 

\section{Colors }

\begin{figure*}[h]
   \includegraphics[trim=0.0cm 4.0cm 0.0cm 4.0cm, clip,width=\linewidth]{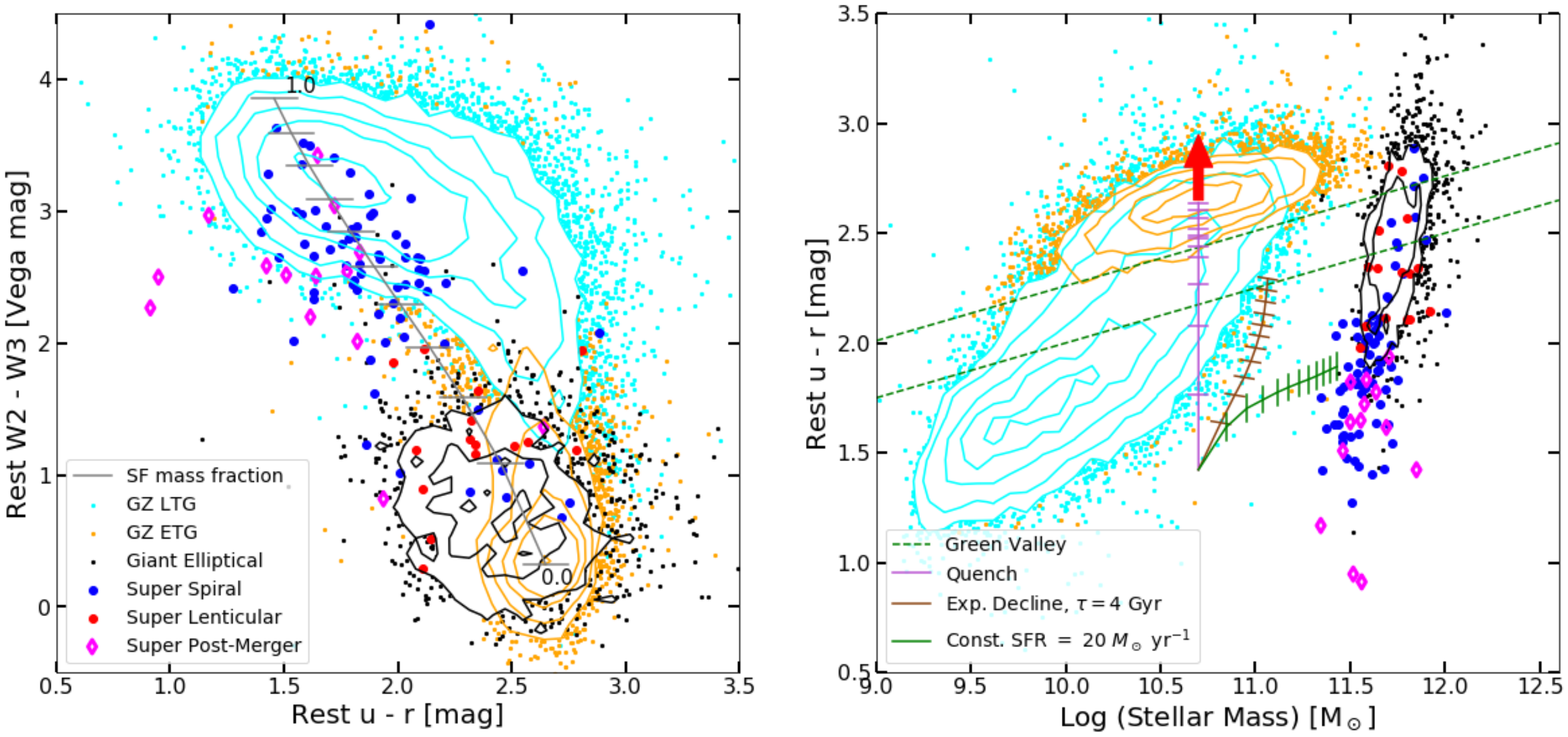}
   \figcaption{
   (a) SDSS and {\it WISE} colors of super spirals (dark blue), super lenticulars (red), super post-mergers (magenta diamonds),   and giant ellipticals  (black) are    
        compared to Galaxy-Zoo (GZ) classified SDSS galaxies at $z=0.02-0.05$  \citep{lss08,sus14, asa14}.  Reddening by Galactic-type dust would increase $u-r$, with no effect on the {\it WISE} color.
        The grey track follows the linear mixing of a young, star-forming stellar population  ($u-r=1.4$, $W2-W3=3.8$) with an old, quiescent stellar population ($u-r=2.65$,  $W2-W3=0.32$). The star forming fraction $f_\mathrm{sf}$ 
        by $W2$ luminosity, which follows mass in stars, is indicated by the tick marks separated by 0.1. The mix with  $f_\mathrm{sf}=0.1$ has $W2-W3=1.09$ and equal contributions to $W3$ luminosity from 
        quiescent and star-forming populations.  
   (b) SDSS color-mass diagram, without any correction for internal extinction. The location of the green valley (green dashed lines), as determined for less massive galaxies by \cite{sus14}, is shown for comparison. 
        We over-plot  evolutionary tracks based on \cite{bc03} stellar population synthesis models, with a starting age of 0.5 Gyr and markers spaced 1 Gyr apart (tick marks at color ages of 1.5, 2.5,...11.5 Gyr). 
        All three galaxies start at $z=6$ with $\log M_\mathrm{stars} = 10.7$, solar metallicity $Z=Z_\odot$, and a stellar population age of 0.5 Gyr, which represents the newly-formed bulge.  In the Quench model (lavender track), 
        the galaxy quenches immediately.  For the other two models, a disk subsequently forms with either constant SFR (green track) or exponentially declining SFR (brown track) with e-folding timescale of  $\tau = 4$ Gyr. 
        Galaxy $u-r$ color gets redder with time even for the constant SFR model as old stellar populations accumulate. The endpoint mass, color, and bulge/disk mass ratio of the constant SFR model is similar 
        to super spirals in our sample.  The red arrow indicates the effect on $u-r$ of increasing the end-point metallicity by 0.2 dex to $1.6 Z_\odot$.
     \label{5}}
\end{figure*}

\begin{figure*}[t]
   \includegraphics[trim=2.5cm 6.0cm 1.0cm 5.0cm, clip, width= 0.5\linewidth, angle=0]{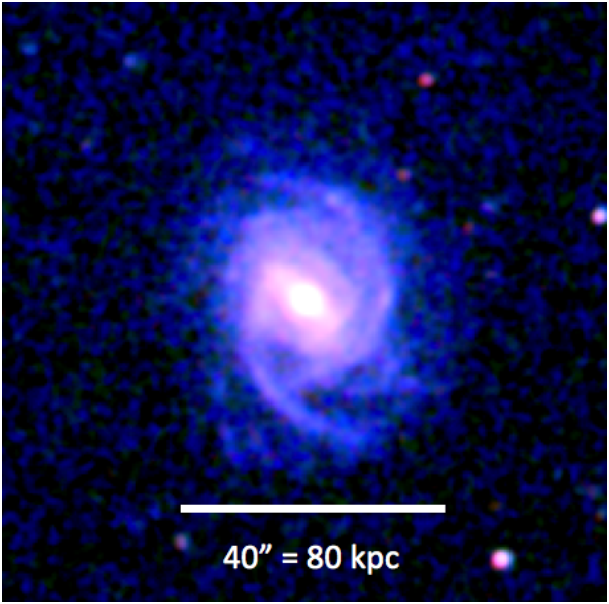}
   \includegraphics[trim=0.0cm 0.0cm 0.0cm 0.0cm, clip, width= 0.5\linewidth, angle=0]{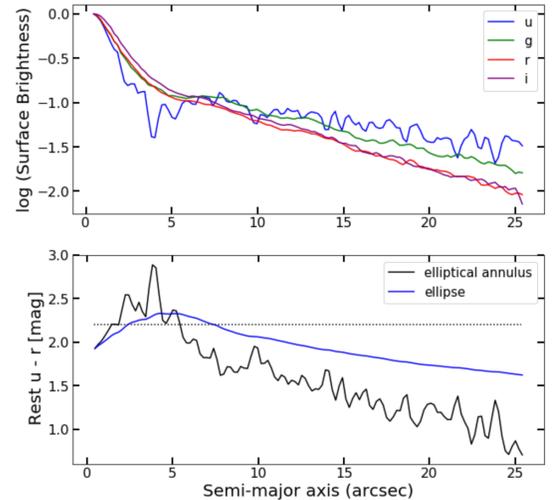}
   \figcaption{Image: super spiral OGC 0543 SDSS $gri$ 3-color image. 
   		    Top plot: Surface brightness profile in SDSS bands, within elliptical annuli, normalized to unity at $r=0\arcsec$. Bottom plot: 
                     $K$-corrected $u-r$ color (solid black line) decreases with radius in the disk, indicating increasing SSFR and decreasing color age or decreasing metallicity. The integrated 
                     $u-r$ color within an elliptical aperture (solid blue line) also becomes bluer as the aperture size is increased. 
                     The optical color separator for the massive blue and red-sequence galaxies in our sample ($u-r=2.2$ mag) is indicated by the dotted line.
    \label{6}}
\end{figure*}

\subsection{Total System Colors}
For lower mass galaxies with $M_\mathrm{stars} = 10^{9}-10^{11} M_\odot$, there is a well-established bimodal distribution in color between blue, star-forming spiral galaxies and red elliptical galaxies with low star-formation rate \citep{sus14}.  Galaxies that fall in the so-called green valley that separates the two populations are primarily disk galaxies (spirals and lenticulars) with relatively low specific star formation rates, plus a small contingent of elliptical
galaxies with modest star formation rates (Fig. 5).  The origin of the bimodal color distribution has been attributed to various star-formation quenching mechanisms, including galaxy mergers and quasar activity.

Most super spirals are blue, with rest frame $u-r<2.2$ mag, similar to less massive Galaxy Zoo late type galaxies \cite[LTGs,][]{lss08}.   The  {\it WISE} $W2 - W3$ color  tracks SSFR, to the extent that it correlates  with $K_\mathrm{s}-W3$,  the basis of our SFR vs. mass plot (Fig. 4).   Super spirals with red $W2-W3$ color have high SSFR, while lenticulars have relatively low SSFR and bluer  $W2-W3$ color.  We find a linear anti-correlation between $W2-W3$ and $u-r$  for our sample of massive galaxies (Fig. 5a),  reflecting an anti-correlation between SSFR and luminosity-weighted mean stellar population age.  Galaxies with the highest SSFR have blue $u-r$ and red $W2-W3$, characteristic of luminous, young stellar populations, while galaxies with lower SSFR have redder $u-r$ and bluer $W2-W3$ color, from a mix of young and old stellar populations. This stands in contrast to lower-mass disk galaxies along the blue sequence, which follow a dog-leg trajectory in color space (Fig. 5a).  The $W2-W3$ color for these galaxies remains red even after they have quenched and moved into the green valley, perhaps because of dust heated by a post-starburst population of A stars \citep{asa14,abl17}.  Super spirals do not appear to follow this trajectory in color space, indicating a different evolutionary history that is consistent with ongoing star formation and a mixture of old and young stellar populations. We can reproduce the observed range of SDSS and WISE colors by a linear mix of a young, star-forming stellar population  ($u-r=1.4$, $W2-W3=3.8$) with an old, quiescent stellar population ($u-r=2.65$,  $W2-W3=0.32$), where we vary the star-forming mass fraction $f_\mathrm{sf}$ (Fig. 5a).  This allows us to estimate the star-forming mass fraction for any given galaxy, and the fraction of light in each band that comes from the two
stellar populations. This model does not apply to quenching galaxies, such as lower-mass lenticulars, which follow a different trajectory in this color space.

While super spirals have similar colors to less massive spirals, they stand dramatically apart in the color-mass plane (Fig. 5b).  Most super spirals have blue $u-r$ colors corresponding to high SSFRs, in spite of their enormous mass in stars.  In order to better understand the stellar populations of super spirals and lenticulars, we created synthetic galaxy colors by summing the stellar population synthesis  (SPS) spectral models of \cite{bc03}.  We assume solar metallicity ($Z=Z_\odot$) for the massive galaxies in our sample. The influence of the mass-metallicity relation in star-forming galaxies \citep[e.g.,][]{t04} on $u - r$ is also examined by increasing the
end-point metallicity by 0.2 dex to $1.6Z_\odot$, the maximum value supported by the SPS models. For a constant or declining SFR, the $u-r $ color reddens monotonically with time, yielding a range of color matching super spirals (Fig. 5b).  In particular, galaxies that form stars at a constant rate become steadily redder with time as old stars accumulate within their disks, reaching $u-r=1.8$ mag after 12 Gyr. Galaxies that are redder than this must have declining star formation rates.  Increasing the metallicity by 0.2 dex to $Z=1.6 Z_\odot$ reddens $u-r$ by 0.25 mag in our SPS models. 

Both super lenticulars and massive ellipticals are on average redder than super spirals, consistent with older stellar population ages.  For a  single stellar population created in a $\delta$-function burst, the $u-r$ color increases from $u-r=1.4$ mag at $t=0.5$ Gyr to $u-r=2.7$ mag at  $t=11.5$ Gyr in our SPS quench model (Fig. 5b).  This type of evolution describes passive galaxies on the red sequence that formed and quenched not long after the big bang. We introduce the color age $t_\mathrm{color}$, appropriate to such a single burst stellar population, in order to characterize the average-luminosity weighted stellar population ages of galaxies. The giant ellipticals have a median $u-r$ color of 2.5 mag, indicating $t_\mathrm{color} \sim 5.5$ Gyr, compared to less massive Galaxy Zoo ETGs that have a median $u-r$ color of 2.7 mag and $t_\mathrm{color} \sim 11.5$ Gyr. The bluer colors and younger color ages of  the massive ellipticals in our  sample may indicate that they are more susceptible to bouts of renewed star formation, perhaps as a result of more frequent mergers or cooling flows in dense environments  \citep{emr06, bhg08}.

\begin{figure*}[t]
   \includegraphics[trim=4.0cm 0.0cm 3.5cm 0.0cm, clip, width=0.5\linewidth, angle=270]{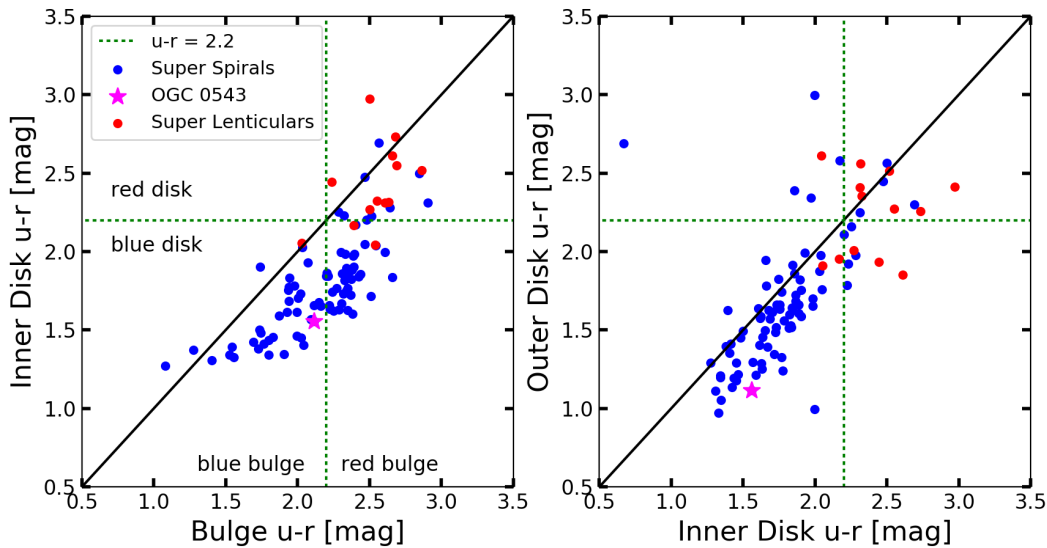}
   \figcaption{  Radial $u-r$ color gradients for super spirals (blue points) and super lenticulars (red points). 
                     (a) The inner disks (middle 1/3 elliptical annulus) of super spirals and super lenticulars are  systematically bluer than their bulges (inner 1/3 ellipse).  (b) The outer disks  (outer 2/3 annulus) 
                      of super spirals are systematically bluer than their inner disks (middle 1/3 elliptical annulus), indicating younger stellar populations.  The diagonal (black line) 
                      delineates equal $u-r$ color.  The horizontal and vertical dotted lines at $u-r = 2.2$ mag separate red and blue stellar populations. The data point for our super spiral case study 
                      subject, OGC 0543 (Fig. 6) is plotted as the star symbol.      
                      \label{7}}
\end{figure*}

\subsection{Color Gradient}

The disks of spiral galaxies typically display a negative color gradient, with bluer colors at larger radii, attributed to a combination decreasing stellar population age and decreasing metallicity with radius 
\citep{dj96,bd00}.  While a gradient in dust extinction can also in principle cause a color gradient, this would require an unrealistically large optical depth and dust scale height \citep{dj96}.

We performed elliptical aperture photometry on the SDSS images of  one of the brightest, nearby super spirals (OGC 0543) to characterize its radial color profile
(Fig. 6).  The integrated, K-corrected color over the full galaxy is $u-r=2.03$ mag, typical for the massive, star-forming galaxies in our sample. The $r$- and $i$-band radial light profiles, which track mass in stars,  are smoother and drop more quickly than the $u$-band profile, which tracks star formation. Oscillations in the $g$-band surface brightness correspond to spiral arms in the stellar disk. The $u-r$ color is bluer in the nucleus than in the inner disk  because of  AGN activity (OGC 0543 has a Seyfert 1 nucleus).  Outside of the nucleus, the $u-r$ color gets progressively bluer with increasing radius in the disk, indicating increasing SSFR and decreasing color age with radius.   The large range in $u-r$  color seen in  the disk of OGC 0543 spans the full range of integrated $u-r$ color for spiral galaxies (Fig. 5b).    The $u$-band surface brightness profile indicates a current star formation rate that declines gradually (by only 0.5 dex) from $6-25\arcsec$ (12-60 kpc).  The exponential $r$-band radial profile over the same interval indicates that a larger surface density of old stars has accumulated in the inner disk over cosmic time, relative to the outer disk.  This is consistent with inside-out growth via gas accretion (possibly accompanied by some inward migration of old stars). 

We measured the surface brightness and color profiles in elliptical annuli at fixed PA for all super spirals and super lenticulars in our sample, to see if they show similar color gradients to OGC 0543.  We summarize our results by comparing the integrated $u-r$ color inside three annuli, with outer semi-major axes of $0.33$, $0.66$, and $1.0$ times the isophotal radius at $r = 25$ mag,  $R_{25}$ (Fig. 7).  We label these three regions bulge, inner disk, and outer disk, though it should be kept in mind that all three regions may contain flux contributions  from the bulge, pseudo-disk, disk, or stellar bar.  

As expected, the inner disks of super spirals are bluer than their bulges (Fig. 7a).  Super lenticulars have systematically redder bulges and disks than super spirals, consistent with older stellar populations and lower SSFR.  We can divide super disk galaxies (spirals and lenticulars) into three categories, based on the $u-r$ colors of their inner disks and  bulges, which can be tied to a range of different star formation histories. Forty-five percent have blue disks and blue bulges, 42\% percent have blue disks and red bulges, and 13\% percent have red disks and red bulges.  The first category corresponds to galaxies with star-forming disks and bulges, the second to star-forming disks and quiescent bulges, and the third to quiescent disks and bulges.  The very bluest bulges (possibly pseudo-bulges) with $u-r<1.7$ mag have color ages of $<1.5$ Gyr, indicating recent star formation. Red disks with $u-r>2.2$ mag and color ages  $>3.5$ Gyr  have quenched star formation, including both lenticulars and red spirals like those found in Galaxy Zoo \citep{bnb09}. 

The outer disks of super spirals are on average 0.2 mag bluer in $u-r$ compared to their inner disks (Fig. 7b).  This is consistent with the color gradient found for OGC 0543 above, confirming a tendency for a younger stellar population age (or lower metallicity) in the outer disks of super spirals.   Although still quite red, the inner disks of super lenticulars are systematically bluer than their bulges, consistent with  younger color ages and later quenching times. Their outer disks show no systematic difference in $u-r$ color with respect to their inner disks. 

\begin{figure*}[t]
   \includegraphics[trim=0.0cm 3.0cm 1.0cm 0.0cm, clip, width=\linewidth]{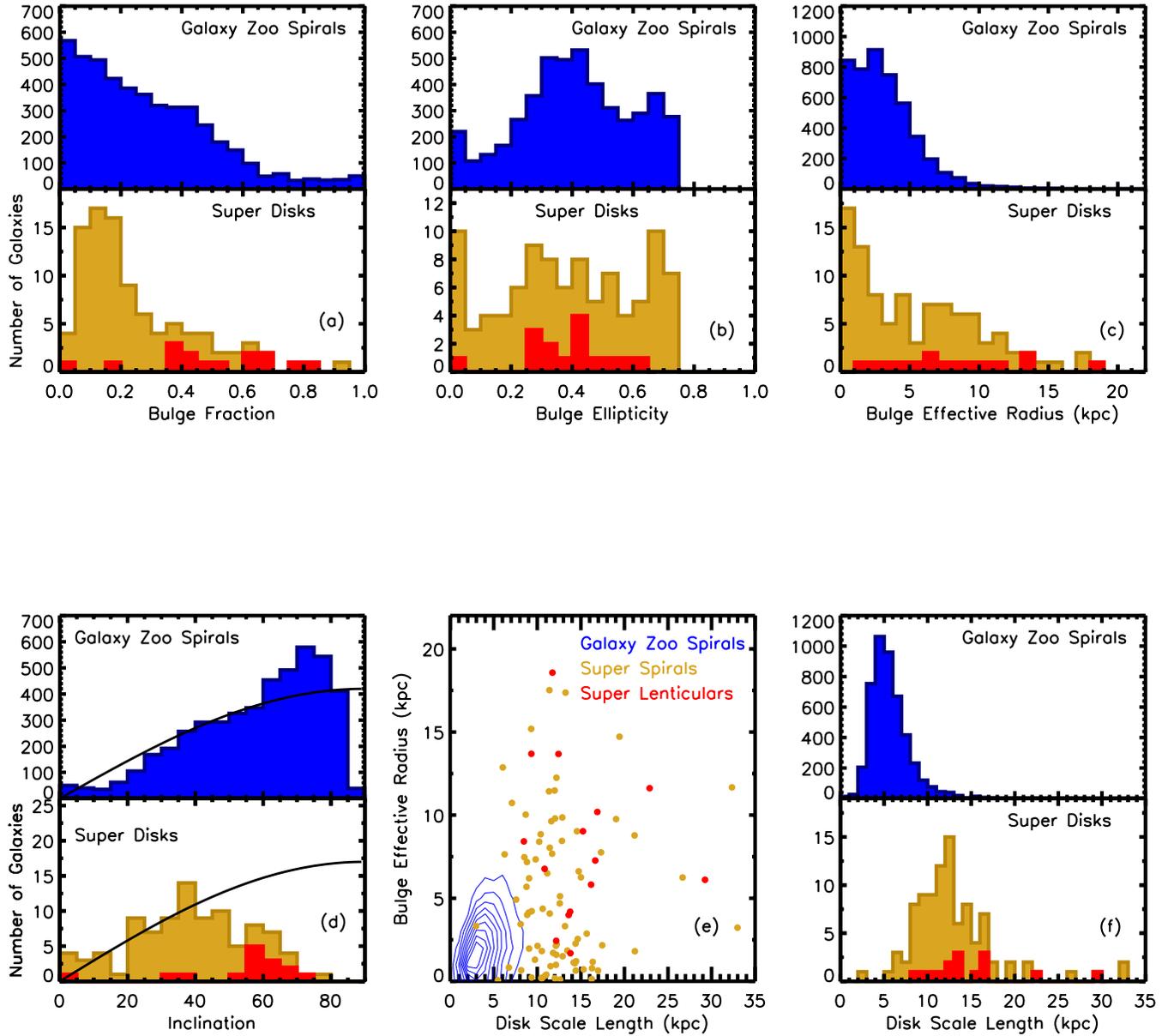}
   \figcaption{Distributions of super spiral (gold histograms) and super lenticular (red histograms) bulge-plus-disk decomposition parameters, as measured by \cite{smp11}, compared to Galaxy Zoo spirals (blue histograms): 
                    (a) bulge to total ($B/T$) $r$-band luminosity fraction, (b) bulge ellipticity $e$, (c) bulge effective radius $R_e$, (d) disk inclination distribution compared to the $\sin(i)$ expectation for randomly oriented disks 
                    (black curve), (e) bulge effective radius vs. disk scale length, and (f) disk exponential scale length.
   \label{8}}
\end{figure*}

\section{Bulge-Disk Decomposition}

We use GIM2D bulge-disk decompositions from \cite{smp11} to compare the quantitative morphologies  and sizes  of super spirals  and super lenticulars to Galaxy Zoo spirals (Fig. 8).  In these decompositions, the bulge is fit by an $n=4$ S\'ersic profile with ellipticity $e$ and effective radius $R_e$, while the disk is fit by an exponential profile with scale length $R_d$ and inclination $i$.  We confirm the results of \cite{oln16} with our larger sample. 

The mean bulge-to-total $r$-band luminosity ratio ($B/T$)  is smaller for super spirals than for Galaxy Zoo spirals (Fig. 8a). The super spiral B/T distribution peaks at $B/T=0.1-0.2$,
while there is no clear peak in the B/T distribution for Galaxy Zoo spirals.  The $B/T$ distribution Galaxy Zoo spirals has a tail with $B/T>0.7$,  where the disk contributes a minor fraction of the $r$-band flux. Only one super spiral (OGC 1514) falls in this part of the distribution.  Otherwise, our visual morphological identification of super spirals is consistent with $B/T$ values indicating a large disk component.  All but two super lenticulars (OGC 0044 and 1386) have $B/T>0.35$, consistent with a major merger origin. 

The distribution of bulge ellipticity (Fig. 8b) is similar for super spirals and Galaxy Zoo spirals. Many super spirals (20$\pm$ 5\%) appear to have bars (Table A5).  While the bar fraction is low compared to the total bar fraction of 65\% for
$L^*$ spiral galaxies at $z=0.14-0.47$ measured with {\it HST} \citep{see08}, it is consistent with the strong bar fraction of 27\% for the same set of galaxies (dropping to 20\% for spirals with the smallest $B/T$).  The lower spatial resolution of SDSS compared to {\it HST} may cause us to miss the weak bars in our sample.  Higher spatial resolution imaging is needed to improve on our measurement of the bar fraction in super spirals. The mean bulge ellipticity for barred super spirals is $<e>=0.50\pm 0.08$, compared to $<e>=0.38\pm 0.02$, for non-barred super spirals. It is likely that the presence of a bar increases the fit bulge ellipticity in some cases.

Both bulges and disks are on average larger in super spirals and super lenticulars than in Galaxy Zoo spirals (Fig. 8c,e,f). The disk scale lengths of super spirals extend to much larger values than those of Galaxy Zoo spirals. The distributions  peak at $R_d=12.5$ kpc and $R_d=4.5$ kpc, respectively.  The distribution of bulge effective radius also extends to much larger values for super disks than for Galaxy Zoo spirals.  However, a significant fraction of super spirals (36/99)  have small, unresolved bulges with $R_e<2$ kpc in spite of their large disk scale lengths ($>5$ kpc).  Both super spirals and super lenticulars cover a large range in $R_e/R_d$, perhaps reflecting a range in merger histories and merger mass ratios.

Even though we dropped the $NUV$-band selection criterion of \cite{oln16}, we still find a large deficit of super spirals at inclinations $i>50\arcdeg$, compared to both Galaxy Zoo spirals and the expected distribution for randomly oriented disks (Fig. 8d).  Roughly 40\% of super spirals must have dropped out of the parent sample because dust extinction in their highly inclined disks caused their $r$-band luminosities to fall below our selection threshold of $L_r> 8 L^*$.   An additional extinction of $\Delta r = 0.6$ mag would suffice to move the brightest face-on super spiral (OGC 0065) below our luminosity selection threshold. The most luminous edge-on spiral galaxy in SDSS I/II at $z<0.3$ is 2MFGC 12344 ($z=0.1407$), with inclination $i=81\arcdeg$.  Its dust lane crosses just above its nucleus, consistent with its high inclination (Fig. 9).  Its apparent luminosity ($L_r=7.9$) is just below our sample selection threshold, and its $r$-band isophotal diameter (120 kpc) rivals the largest super spiral in our sample (OGC 0139: $D=134$ kpc, $L_r= 13.4$).  Selection by NIR luminosity may help to recover many more of these edge-on, dust-obscured super spirals.  Super lenticulars are preferentially selected at intermediate inclinations of $50-75\arcdeg$, because face-on lenticulars are difficult to distinguish from ellipticals, and because edge-on lenticulars may be misclassified as spirals.  We attribute the similar excess of Galaxy Zoo spirals at inclinations of $60-80\arcdeg$ to a population of lenticulars.

\begin{figure}
   \includegraphics[trim=1.0cm 4.0cm 1.0cm 4.0cm, clip, width=0.95\linewidth]{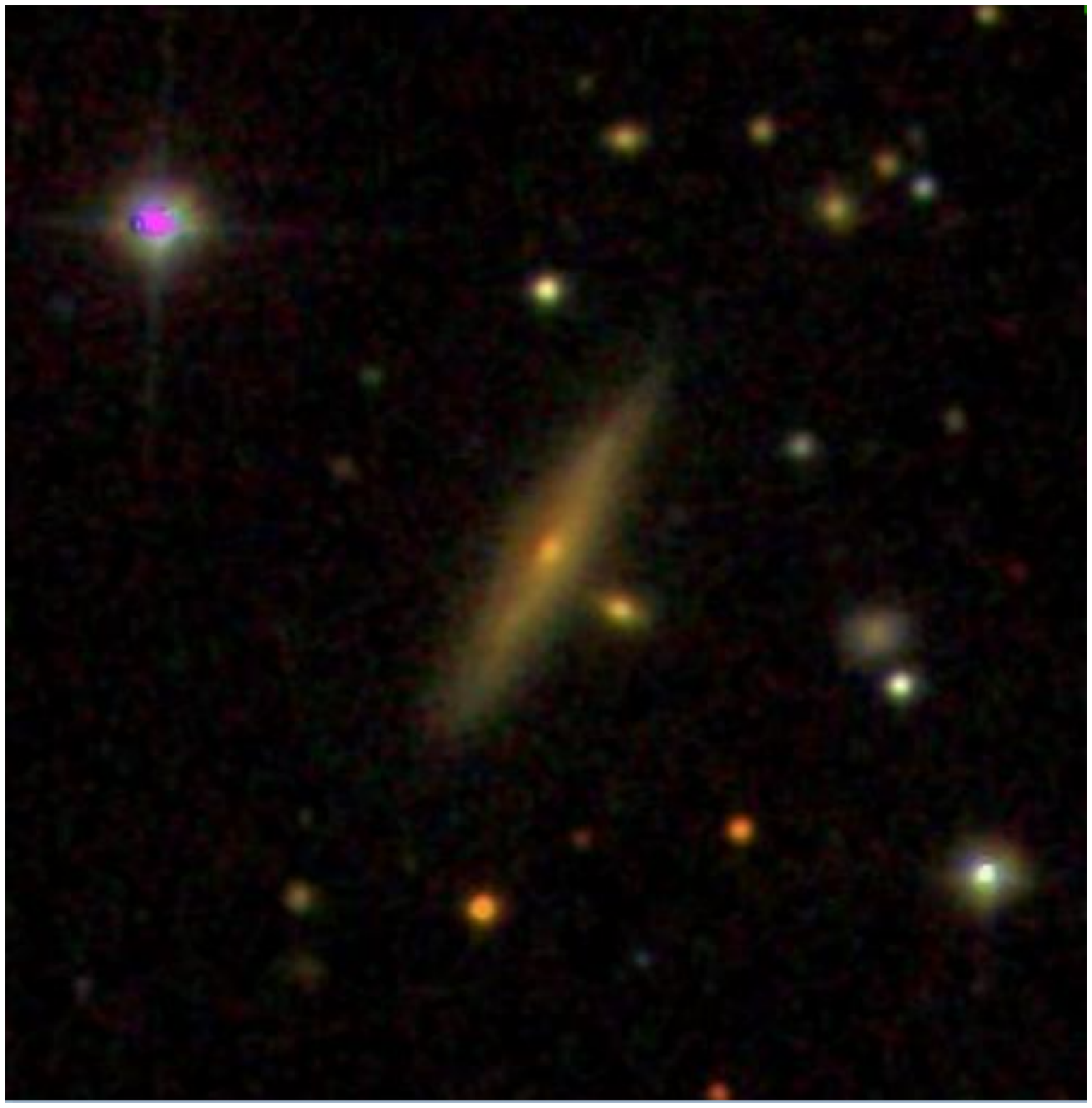}
   \figcaption{SDSS image of 2MFGC 12344, the most luminous nearly-edge-on spiral galaxy in SDSS I/II. The field of view is 250 kpc on a side.
    \label{9}}
\end{figure}

\section{Environment}

Most super spirals and super lenticulars are found in moderately dense environments (Figs. A4 \& A5), with an average of  $7.8 \pm 0.3$ SDSS galaxies within a projected radius of 150 kpc. In comparison,  OGC giant ellipticals have 
on average roughly twice as many galaxies within the same radius ($14.5 \pm 0.1$).  We find only 8 super spirals in dense environments with 14 or more apparent companions within a projected radius of 150 kpc: OGC 0299, 0586, 0799, 1023, 1304, 1329, 1457, and 1559). While most companions are more than 1 magnitude fainter than the super spiral, some are potentially massive enough to result in a major, disruptive merger (e.g., companions to OGC 0299, 0799, 1304, and 1457). 

We searched for known galaxy clusters and groups within $2\arcmin$ of each super spiral, using NED. We find that 28\% of super spirals and super lenticulars appear to be associated with clusters or
groups of galaxies (Table A7).  For these, we used the NED Environment Search tool to count the number of galaxies with separations and redshifts that put them 
within 1 Mpc and 500 km s$^{-1}$ ($N1$) or within 10 Mpc and 5000 km s$^{-1}$ ($N10$).  While these numbers give a rough sense of cluster richness, they must be quite incomplete for
galaxies at the highest redshifts.  Most cluster or group members would not be luminous enough to make it into the SDSS spectroscopic sample.  Indeed, it is seen that clusters associated
with  the lowest redshift super spirals have the most SDSS redshifts (e.g., 310 for OGC 1559 at $z=0.186$), while candidate clusters associated with super spirals at the highest redshifts 
have fewer SDSS redshifts (e.g. 23 for OGC 044 at $z=0.293$).  Deeper redshift surveys are necessary to measure the richness and velocity dispersions of the highest-redshift candidate
clusters.

For four super spirals in the richest clusters (OGC 0345, 0516, 1268, and 1304), we generated velocity plots with NED Environment Search (e.g., Fig. 10).  The locations of the super spirals within the velocity distributions confirm that they are either cluster or supercluster members.  The large velocity dispersions of these clusters (full range of $\pm 3000$ km s$^{-1}$) indicate that they are very massive.  However, no super spirals reside at cluster centers, and their velocities relative to the mean cluster redshift are large ($\sim 2000$ km s$^{-1}$).  This indicates that these super spirals did not form at the cluster center of mass, but rather in the outskirts.  This is confirmed for four
super spiral BCGs (OGC 0170, 0345, 1268,  and 1304) recently observed with {\it XMM-Newton} by \cite{blk18}, which turn out to be located at large distances (150-1100 kpc)  from the X-ray centroids of their host clusters.
However, one super lenticular that they observed (OGC 0073) does turn out to be located at the center of a low-mass ($M_{500}=10^{14} M_\odot$) galaxy cluster.

The presence of some super spirals in clusters is consistent with their high mass in stars and likely high halo mass.  Super spirals may form preferentially in regions of the universe with relatively high overdensities, where a lot of gas is available to accrete onto their dark matter halos. On the other hand, a location at the periphery rather than the center of clusters is consistent with the morphology-density relation.  We would not expect super spiral galaxies to survive as long as they have if they were at the centers of massive clusters, where they would be subject to frequent harassment by cluster galaxies.  

\begin{figure*}
   \includegraphics[trim=0.0cm 0.0cm 0.0cm 0.0cm, clip, width=0.6\linewidth]{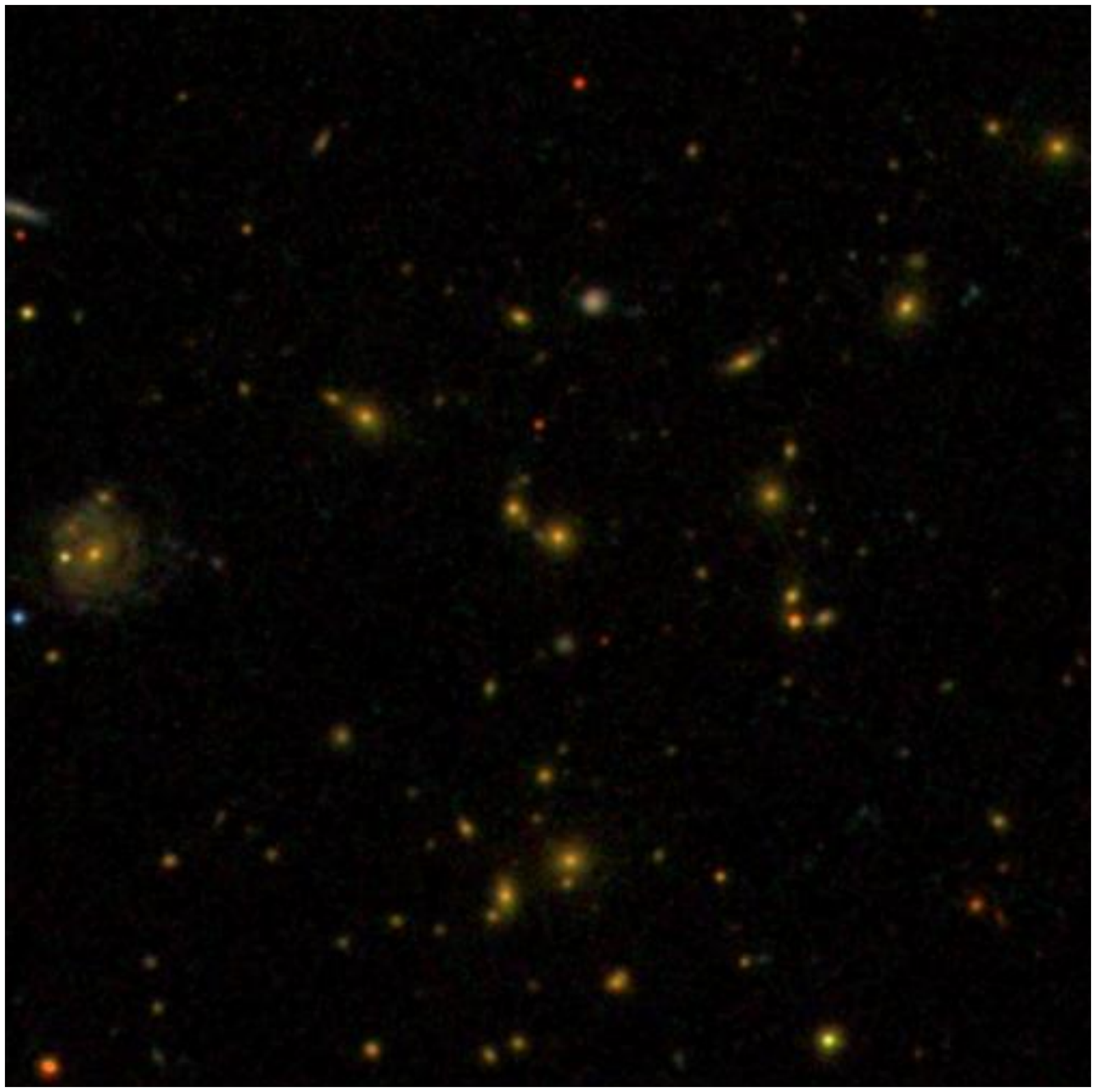}
   \includegraphics[trim=0.0cm 0.0cm 0.0cm 0.0cm, clip, width=0.5\linewidth]{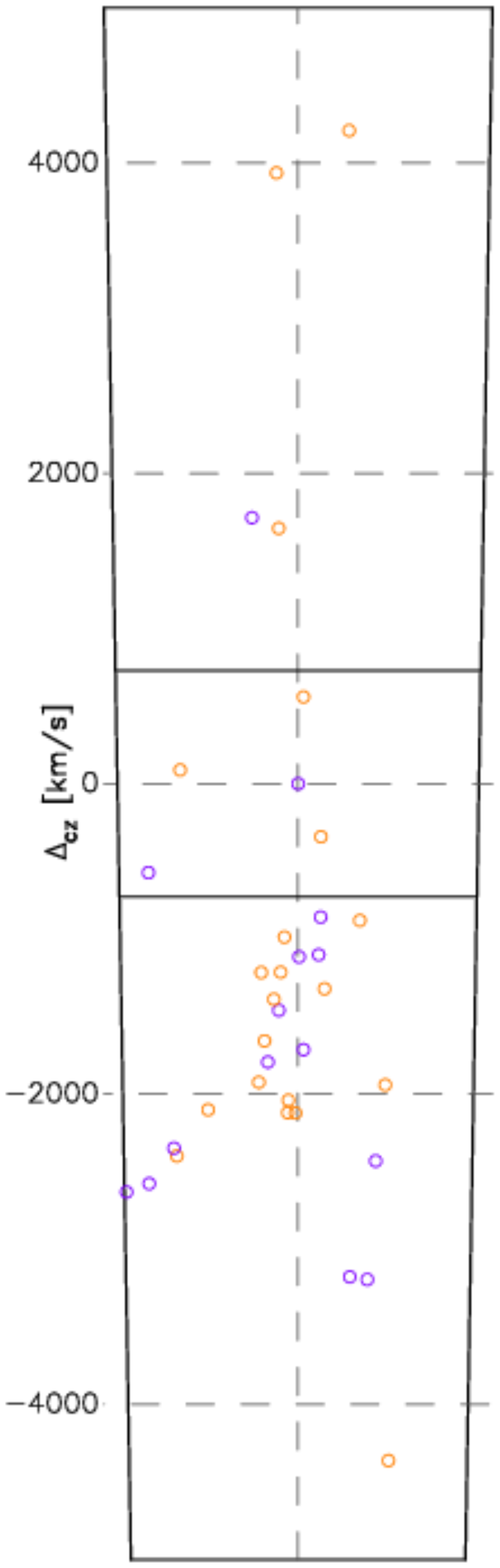}
   \figcaption{ Left:  Super spiral OGC 0345 (at left edge) in galaxy cluster WHL J092608.1+240524.  SDSS image is 200 arcsec = 716 kpc on a side. Right: Position-velocity diagram from NED environment search, 10 Mpc wide and centered on OGC 0345. Orange and purple correspond to galaxies above and below the plane of OGC 0345, respectively.
    \label{10}}
\end{figure*}

\section{Galaxy Mergers}

Many super spirals and super lenticulars (41\%) appear to be involved in mergers (Table A8).  There are 14 candidate interacting pairs or triples with projected separations of 28--131 kpc and 27 advanced mergers with either double nuclei or overlapping disks with projected separations between 4 and 32 kpc. Only three of these pairs are spectroscopically confirmed (OGC 0044, 0299, and 0984). The remaining double nuclei, double disks, and interacting companions need to be verified spectroscopically to rule out chance alignments, but their appearance is quite suggestive of dynamical interactions.  Many of them are highly distorted or have tidal tails or debris that confirm their merger status.  We find 7 candidate collisional ring galaxies (Table A8) and 2 galaxies with asymmetric, $\nu$-shaped arms (OGC 0290 and 1423).   We find four additional galaxies with three or more arms that we do not classify as mergers (OGC 0256, 0926, 1046, and 1323), which may also have suffered a recent dynamical disturbance. 

A high merger percentage is expected for galaxies as massive as super spirals and super lenticulars. \cite{hbc10} estimate that galaxies with $M_\mathrm{stars}=10^{12} M_\odot$ experienced  on average 1.5 major mergers with mass ratio $\mu>0.3$ and 5.5 minor mergers with $0.1< \mu <0.3$ since $z=2$, based on halo occupation statistics.  Most mergers experienced at late times by such massive galaxies are minor mergers because equally massive companions are quite rare. The fraction of massive galaxies with currently visible ongoing mergers can be estimated from the merger rates predicted by their model, multiplied by the merger visibility timescale.  They predict a $\mu >0.1$ merger rate of  0.2 Gyr$^{-1}$ at $M_\mathrm{stars}=10^{11} M_\odot$, increasing to 2 Gyr$^{-1}$ for $10^{12} M_\odot$ galaxies at $z=0.3$.  For a merger visibility timescale of 1.0 Gyr, the corresponding merger fraction ranges from 0.2-1.0.  This is roughly consistent with the observed super spiral merger fraction of 0.41 and merger rate of  0.4 Gyr$^{-1}$, for a median galaxy mass of $10^{11.6} M_\odot$.  If the 11/14 {\it star-forming} super post-mergers in our sample are also the product of super spiral major mergers, this gives a destructive major-merger fraction of 0.12 and rate of 0.12 Gyr$^{-1}$, assuming a post-merger settling timescale of 1.0 Gyr \citep{ljc08}.  During this settling time, obvious tidal signatures of the merger will disappear, and the $u-r$ color of the post-merger $u-r$ color will redden by 0.35 mag to $u-r$=1.75 mag (Fig. 5), at which point it will be difficult to distinguish from other quenched giant ellipticals in our sample. Including 38 ongoing super spiral mergers and 11 star-forming super post-mergers and excluding super lenticular mergers yields an overall  star-forming merger fraction of 0.52 and merger rate of  0.5 Gyr$^{-1}$.

The ability of super spirals to survive under such a high merger rate depends on whether the mergers are major mergers or minor mergers and also on the gas fraction \citep{hcy09}.  We find 8 super spirals in major pairs 
($\Delta g <1.19$ mag) with separations $<150$ kpc (Table 8), that are likely to undergo transformative major mergers in the next Gyr.   The remaining 35 mergers are minor pairs or double nuclei galaxies whose disks survived 
(including 3 S0/Sa galaxies).  It is difficult to estimate the progenitor mass ratio for the double nuclei galaxies, but they must have either been minor mergers or unusually gas-rich mergers in order to retain or re-form their disks.  If they are minor mergers, then the ratio of major to minor mergers is 0.18, roughly consistent with the 0.35-0.1 major/minor merger ratio predicted by \cite{hbc10}  for galaxies with $M_\mathrm{stars}=10^{11-12} M_\odot$.  It will be necessary to obtain H {\sc I} and CO emission line measurements, or alternatively sub-mm dust continuum measurements, in order to determine the gas fractions of super spirals and its impact on their merger survivability.

\section{Discussion: Super Spiral Growth and Destruction}

Armed with an estimate of the merger rate (Section 7) and SSFRs (Section 3), measured at a median redshift of $z=0.22$, we can predict the space densities and masses in stars of super spiral progenitors at 
intermediate redshift ($z=1.0$). Over that redshift interval (a period of 5.2 Gyr), both rates should increase by roughly a factor of 2, according the model of \cite{hbc10}, yielding an average merger rate of 
1.3 Gyr$^{-1}$  and 1/SSFR$=20.6$ Gyr along the ridge line of the SFMS. At this merger rate, super spirals will on average undergo 5.5 minor mergers  with  $\mu \sim 0.1 $ and 1.2 major mergers with $\mu>0.3$ between $z=1.0$ and $z=0.22$.  Roughly 70\% of super spirals will be transformed into giant ellipticals or super lenticulars by major mergers over this redshift interval, reducing the comoving space density of super spirals 
by a factor of 3.3 and increasing the comoving space density of giant ellipticals by a factor of 1.3.

Typical super spiral progenitors will increase their disk mass in stars by 33\% from $z=1.0$ to the present via steady, in-situ star formation.   Super spirals that do not experience a major merger in this time will increase their mass in stars by another 69\% from direct acquisition of stars through minor mergers, most of which will be incorporated into their halos and thick disks.   Gas-rich minor mergers will also add gas to super spiral disks, which may be transformed into additional mass in stars via  merger-induced starbursts.  Assuming  a gas fraction of 0.5 for the secondary galaxy, and that half of this gas is converted into stars,  super spiral mass in stars will grow by an additional 35\% from merger induced starbursts, with most of this mass incorporated into the disk. Together with steady in situ star formation, this yields a combined 68\% gain in disk mass in stars.

 From this model, we predict that 70\% of super spirals were destroyed by major mergers from $z=1.0$ to $z=0.2$ and transformed into giant ellipticals or super lenticulars.  The remaining 30\% that survived through this time period increased their total mass in stars by a factor of 2.4, with half going to the disk and half going to the halo,  while maintaining a relatively low $B/T$ mass ratio.  The answer to the question of how super spirals survive is twofold.  First, the ones that did survive are a factor of two less massive than giant ellipticals and therefore reside in regions that are on average a factor of 2 less dense, resulting in a lower overall merger rate.  Second, their large masses protect them from mergers, such that 82\% of all super spiral mergers  at $z=0.2$ are minor mergers that do not destroy their disks. 

\section{Conclusions}

We present a catalog of 84 SDSS super spirals, 15 super lenticulars, 14 post-mergers, and 1400 giant ellipticals, selected for  $r$-band luminosity $L_r>8L^*$ and redshift $z<0.3$.  These galaxies represent the most massive galaxies in their redshift range, with masses in stars of $10^{11.3}-10^{12.3} M_\odot$. Super spirals are characterized by very large, high-mass, high-surface brightness, actively star-forming disks that fall on or below the star-forming main sequence of galaxies.  Super lenticulars have low SSFRs, particularly red  optical colors,  and no discernible spiral arms. Super post-mergers may be the product of super spiral major mergers, caught during the quenching phase, before they have completely ceased star formation.

The location of super spirals in {\it WISE}-SDSS color space shows that their star-forming disks contain a mix of young and old stellar populations. Their {\it WISE} [4.6]-[12] colors are relatively blue compared to less-massive spirals because they have  accumulated large masses of old stars in their disks,  resulting from early formation in some of the most massive dark matter halos.  Super spiral disks are red on the inside and blue on the outside, consistent with ongoing growth and inside-out formation by accretion of cold gas and minor mergers. Super spirals must form stars at a high rate throughout their lifetimes in order to grow their massive, gigantic disks and maintain their blue integrated colors.

Super disk galaxies are primarily found in moderate density environments, with on average half as many companions within 150 kpc, compared to giant ellipticals.  For the 28\% found in galaxy clusters, they are located at the cluster outskirts,  with high relative velocities ($\sim 2000$ km s$^{-1}$), consistent with the morphology-density relation.  Star formation quenching is not an inevitable conclusion for the most massive spiral galaxies, provided that  they do not live at the dense centers of the largest galaxy clusters.  Super spirals that  do suffer major mergers may be transformed into super lenticulars or giant elliptical galaxies, providing a possible pathway to generate isolated giant elliptical galaxies outside of galaxy clusters.

A large percentage (41\%) of super disk galaxies are involved in ongoing mergers or interactions with other galaxies.  We suggest that some super spirals survive because most mergers are minor mergers for such massive galaxies. Super spirals have low bulge/total luminosity ratios, also consistent with disk building by cold gas accretion and bulge construction by minor mergers. While a large reservoir of high-angular momentum gas could also aid in preserving the structure of super spirals, it will require sensitive radio and sub-mm observations to  establish whether or not this is actually the case.

Super spirals and super lenticulars are disk galaxy counterparts to the most massive, giant elliptical galaxies. Star formation remains unquenched in most super spirals, in spite of their very large masses of old stellar populations. In fact, they appear to have survived through the ages in moderately dense environs, by virtue of their large masses.  Anticipating future studies, the extreme masses, luminosities, and sizes of super spirals will open new parameter space for testing galaxy scaling laws and theories of massive galaxy formation and evolution. The large ongoing merger fraction and variety of merger mass ratios and geometries present in super spiral systems also provide a unique opportunity to study the impact of mergers on massive spiral galaxy structure, star formation,  and evolution. 

\appendix{}
\renewcommand{\thefigure}{A\arabic{figure}}
\setcounter{figure}{0}
\renewcommand{\thetable}{A\arabic{table}}
\setcounter{table}{0}

\subsection{Catalog of the Most Optically Luminous Galaxies}
The Ogle et al. (2018)  Galaxy Catalog (OGC) is presented in this Appendix (Table A1), including catalog names, SDSS $r$-band magnitude, K-corrected $r$-band luminosity, redshift, morphology, and spectral type. 
Catalog sources that we reject based on photometric contamination, bad redshifts, and spatial overlap with foreground or background sources are presented in Tables A2-A4.  Derived physical properties are presented for super spirals,
lenticulars, and post mergers in Table A5.  We present images of  super spirals, lenticulars, and post-mergers  at a scale of 150 kpc on a side  in Figures A1-A3 and images of their larger (300 kpc $\times$ 300 kpc) environs  in Figures  A4-A5.
Super spiral and lenticular associations with galaxy clusters and groups are given in Table A7,  including cluster names, redshifts, galaxy counts, and separation. Candidate super spiral and lenticular mergers and pair separations are found in
Table A8.

\subsection{Non-spiral AGN and QSO hosts}

There are 12 non-spiral galaxies with SEDs contaminated by bright type-1 AGNs or stellar objects that we excluded from our analysis (Table A6 and Fig. A3). Among these are  6 spectroscpically verified QSO host galaxies (OGC 0239, 0302, 0377, 0889, 1239, and 1245) and 3 known BL Lac hosts (OGC 0615, 0962, and 1229), where the AGN may contribute significantly to the $r$-band luminosity. There are another 3 galaxies accompanied by bright stellar objects (OGC 0307, 0469, and 0646) which may be either stars or QSOs, for which no SDSS spectra are available.  We originally identified QSO host OGC 0302 as a super spiral \citep{oln16}, but HST imaging shows it to be a disturbed, possibly post-merger galaxy rather than a spiral galaxy (Fig. A6).

\subsection{Gravitational Lenses}

Galaxy-scale gravitational lenses are potential contaminants to our sample that may artificially boost the $r$-band flux or present arc-like features that can be mistaken for spiral arms (Fig. A6). For example, the giant elliptical galaxy OGC 0203 (2MASX J11125450 $+$1326093 in A1201)  is a known gravitational lens, with a lens arc projected $2\farcs0$ from its center \citep{ess03}.  We initially identified the previously unknown gravitational lens OGC 0200 (2MASX J08355126$+$3926220) as a super spiral, but an existing HST image  shows multiple lens arcs that masquerade as faint spiral arms in the lower resolution SDSS images. Finally we identify the brightest cluster galaxy OGC 1565 (2MASX J21531028$+$1154551, z=0.289) as a gravitational lens candidate and possible Einstein ring by the unusual red ring that encircles it (Fig. A6).

\begin{figure*}
   \includegraphics[trim=2.0cm 4.0cm 2.0cm 3.0cm, clip, width=\linewidth]{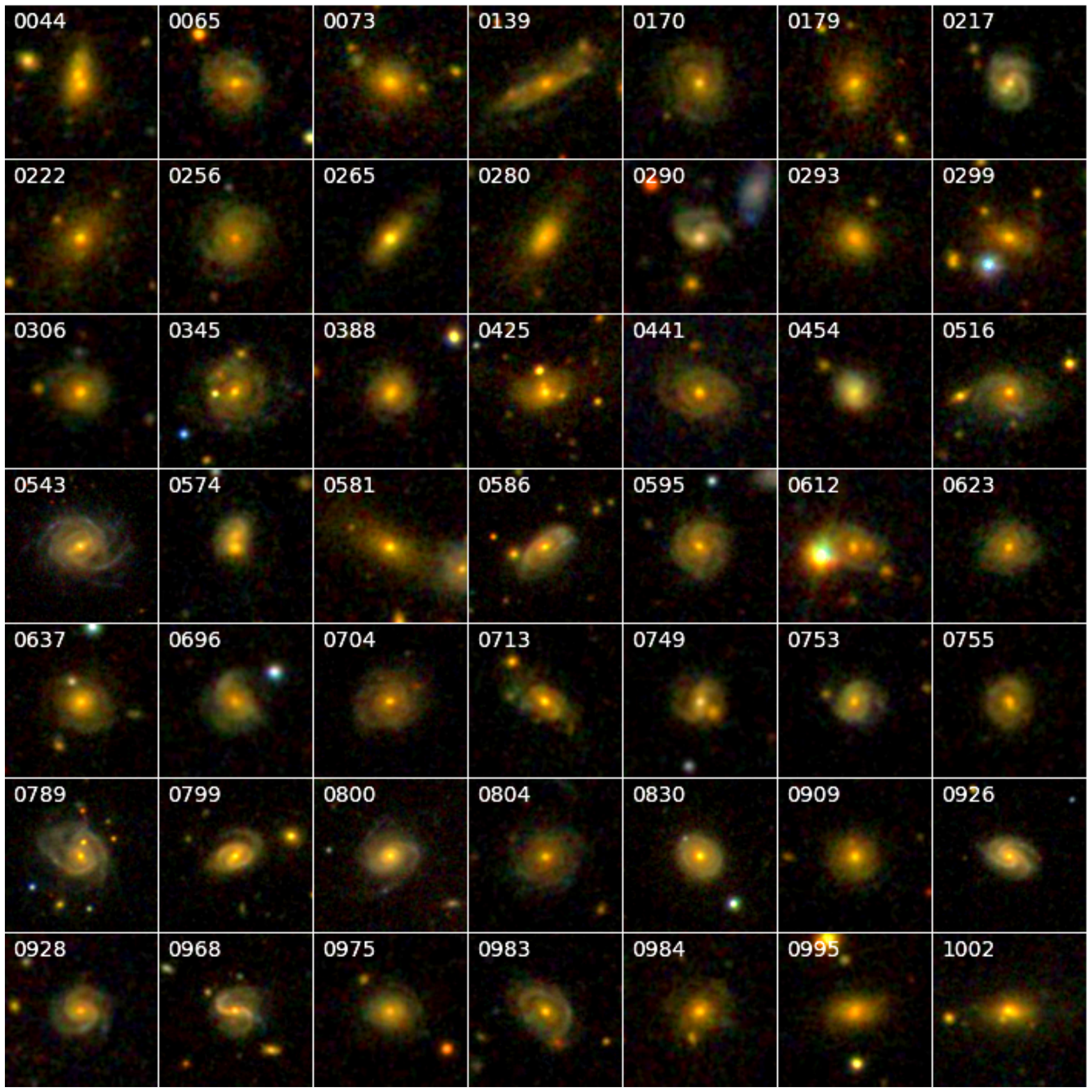}
   \figcaption{SDSS images of super spiral and super lenticular galaxies OGC 0044 - 1002, ordered by decreasing $r$-band luminosity.  Each image is 150 kpc on a side.
    \label{A1}}
\end{figure*}
 
\begin{figure*}
   \includegraphics[trim=2.0cm 4.0cm 2.0cm 3.0cm, clip, width=\linewidth]{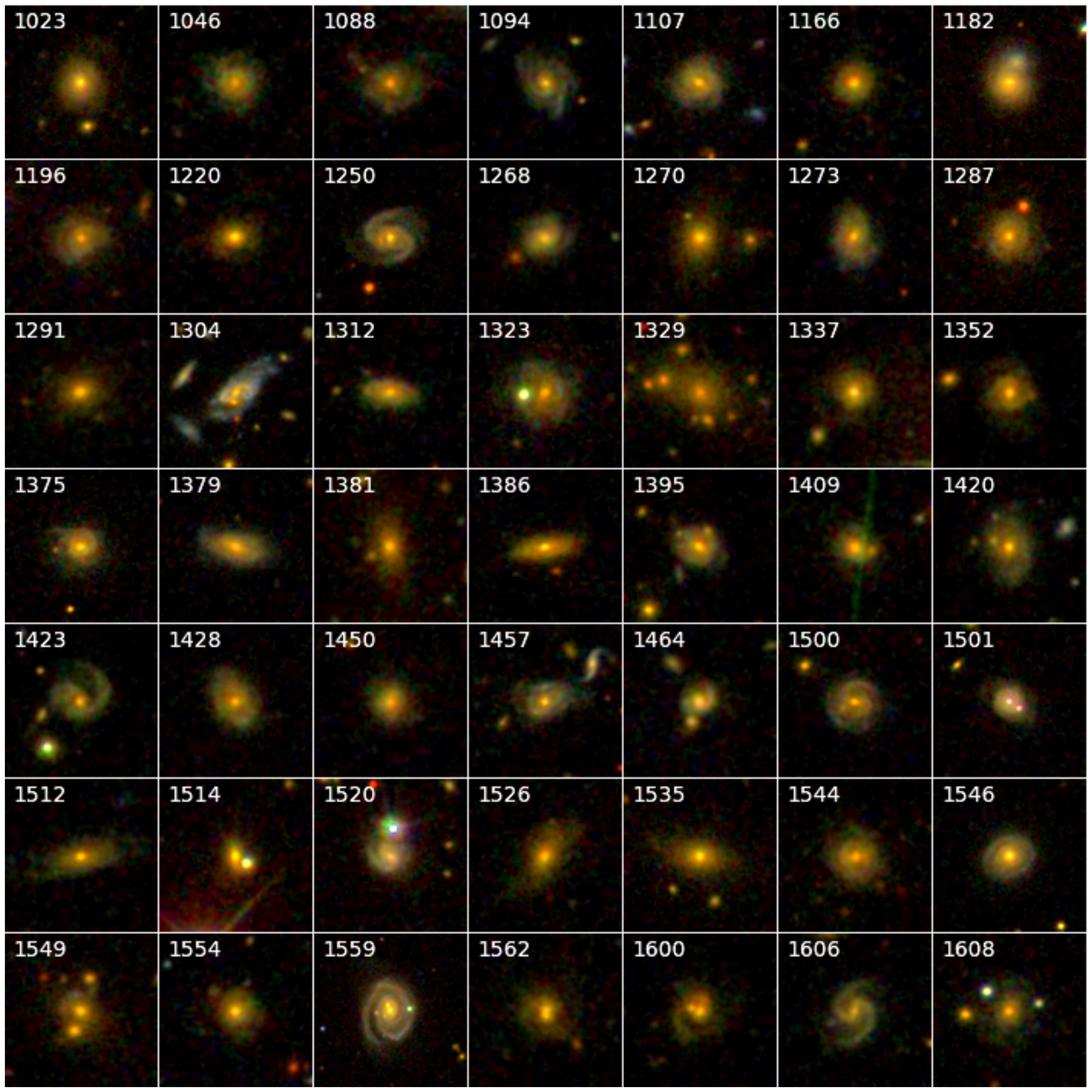}
   \figcaption{SDSS images of super spiral and super lenticular galaxies OGC 1023 - 1608, ordered by decreasing $r$-band luminosity.  Each image is 150 kpc on a side.
    \label{A2}}
\end{figure*}

\begin{figure*}
   \includegraphics[trim=3.0cm 7.0cm 3.0cm 5.0cm, clip, width=\linewidth]{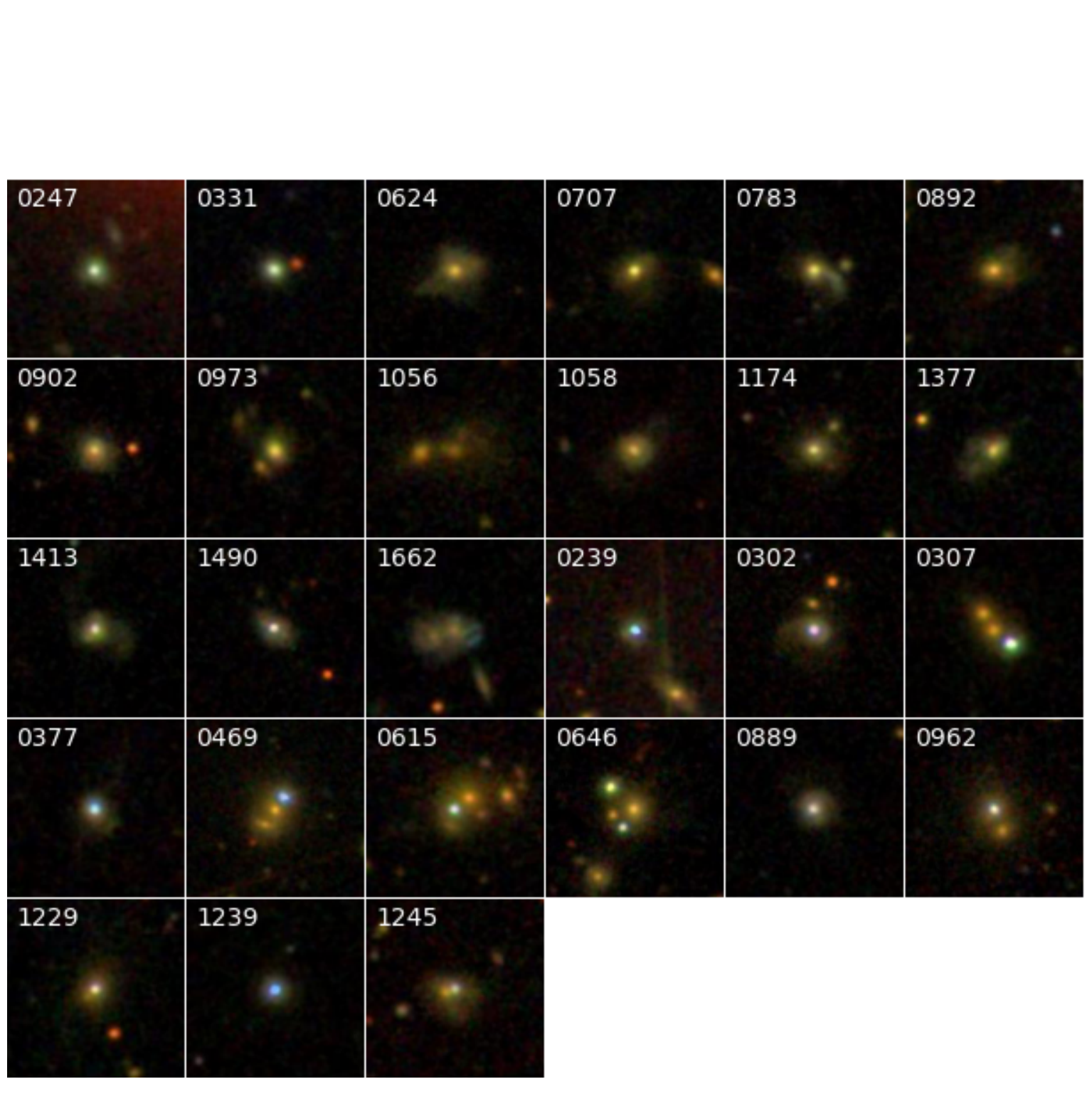}
   \figcaption{SDSS images (150 kpc on a side) of  super post-mergers (OGC 0247-1662), followed by
                     non-spiral QSO or BL Lac hosts or galaxies with compact companions (OGC 0239-1245).
                     \label{A3}}                  
\end{figure*}

\begin{figure*}
   \includegraphics[trim=1.7cm 4.0cm 1.7cm 4.0cm, clip, width=\linewidth]{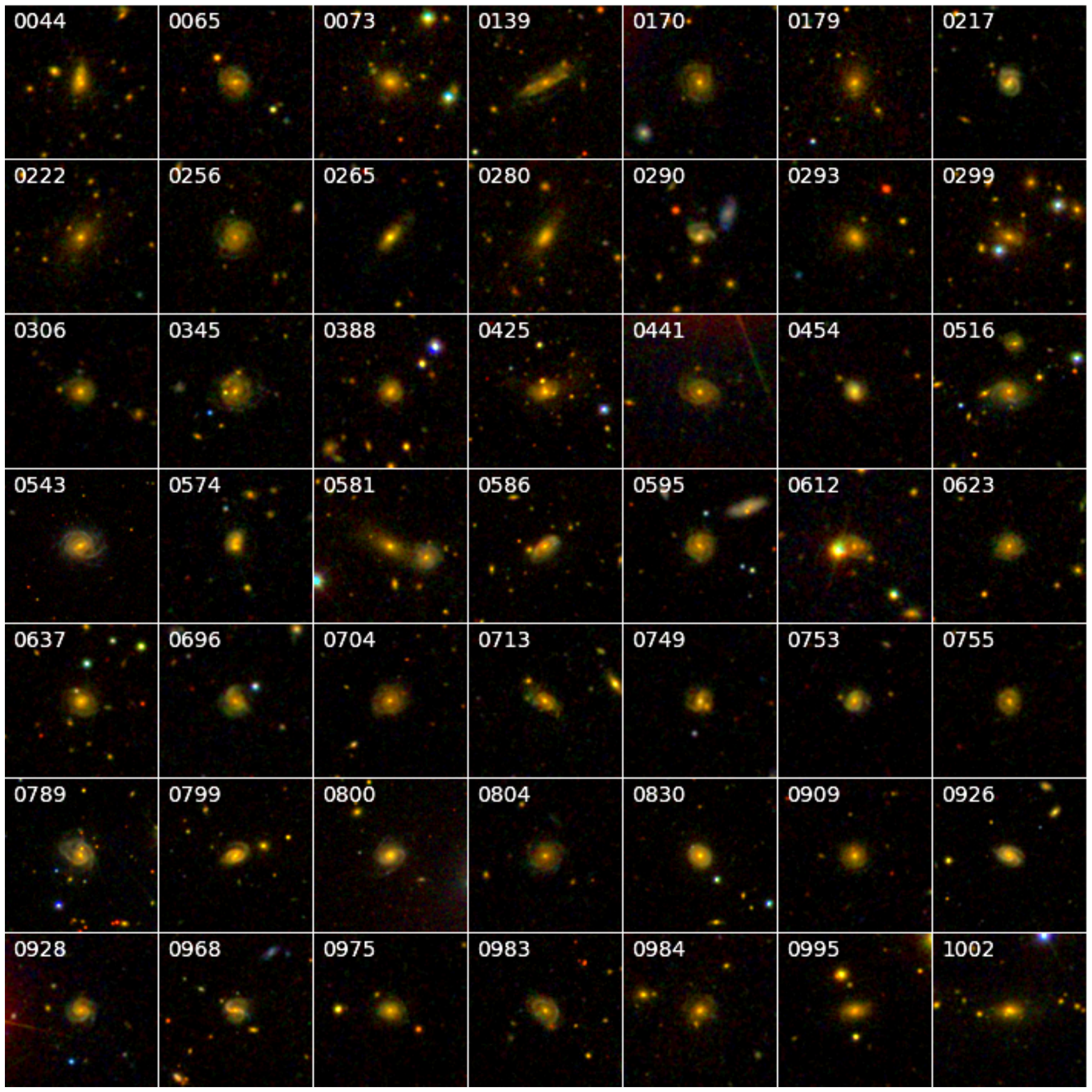}
   \figcaption{SDSS images of the environs of super spiral and super lenticular galaxies OGC 0044 - 1002, ordered by decreasing $r$-band luminosity.  Each field-of-view is 300 kpc on a side.
   \label{A4}}
\end{figure*}

\begin{figure*}
   \includegraphics[trim=1.7cm 4.0cm 1.9cm 4.0cm, clip, width=\linewidth]{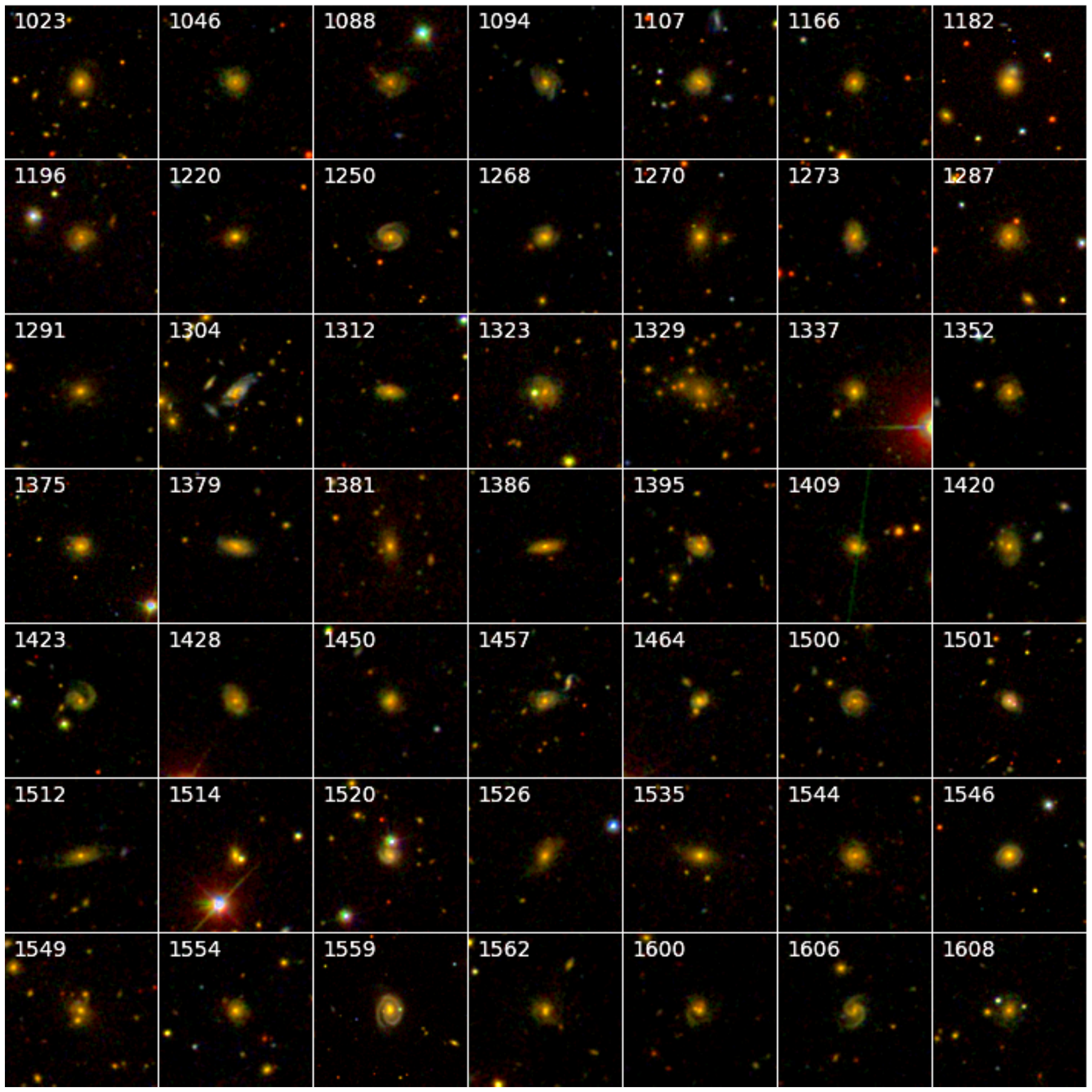}
   \figcaption{SDSS images of the environs of super spiral and super lenticular galaxies OGC 1023 - 1608, ordered by decreasing $r$-band luminosity.  Each field-of-view is 300 kpc on a side.
    \label{A5}}
\end{figure*}

\begin{figure}
   \includegraphics[trim=1.5cm 0.2cm 1.8cm 0.0cm, clip, width=0.363\linewidth]{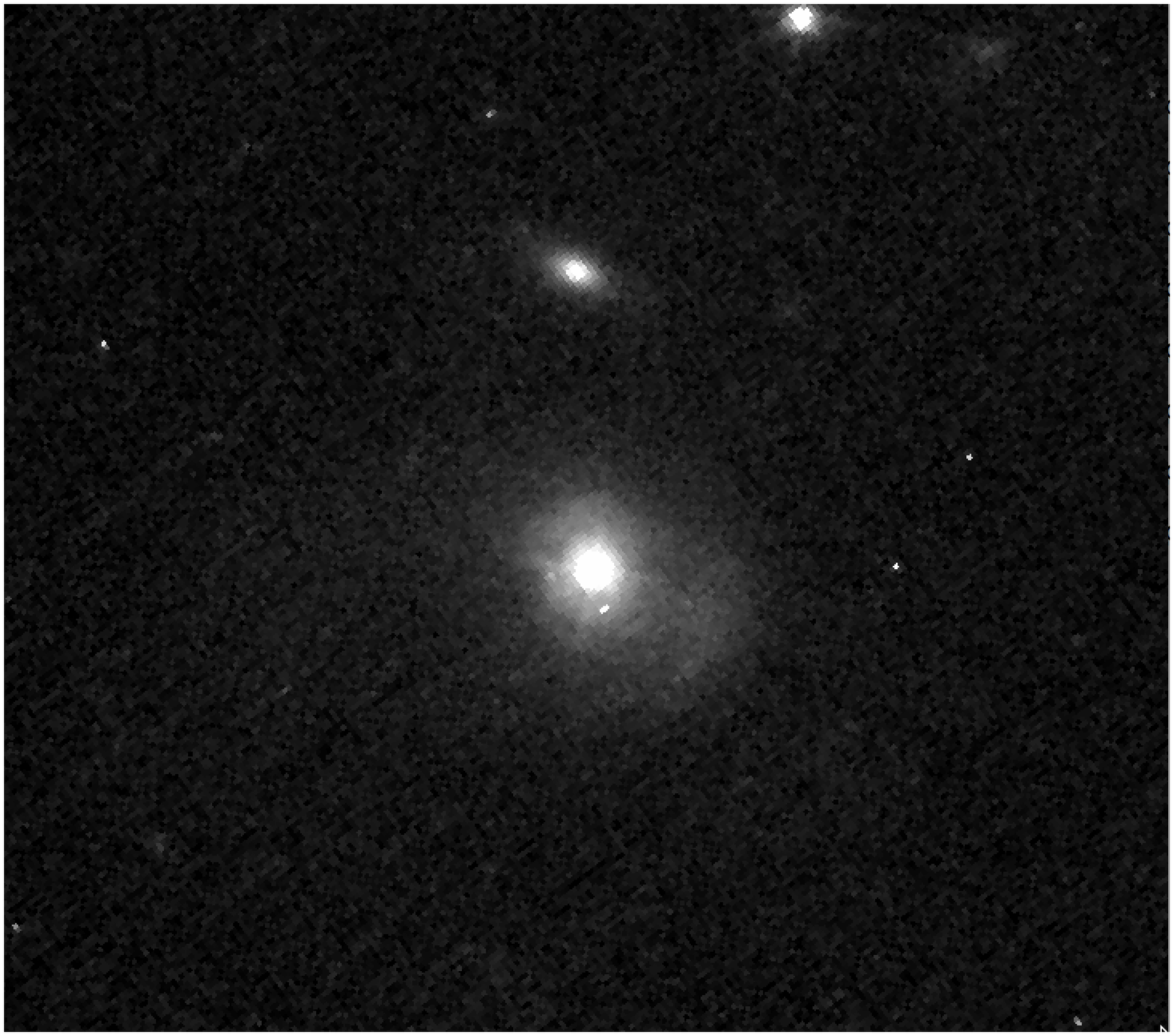}
   \includegraphics[trim=3.5cm 6.6cm 3.5cm 4.2 cm, clip, width=0.31\linewidth]{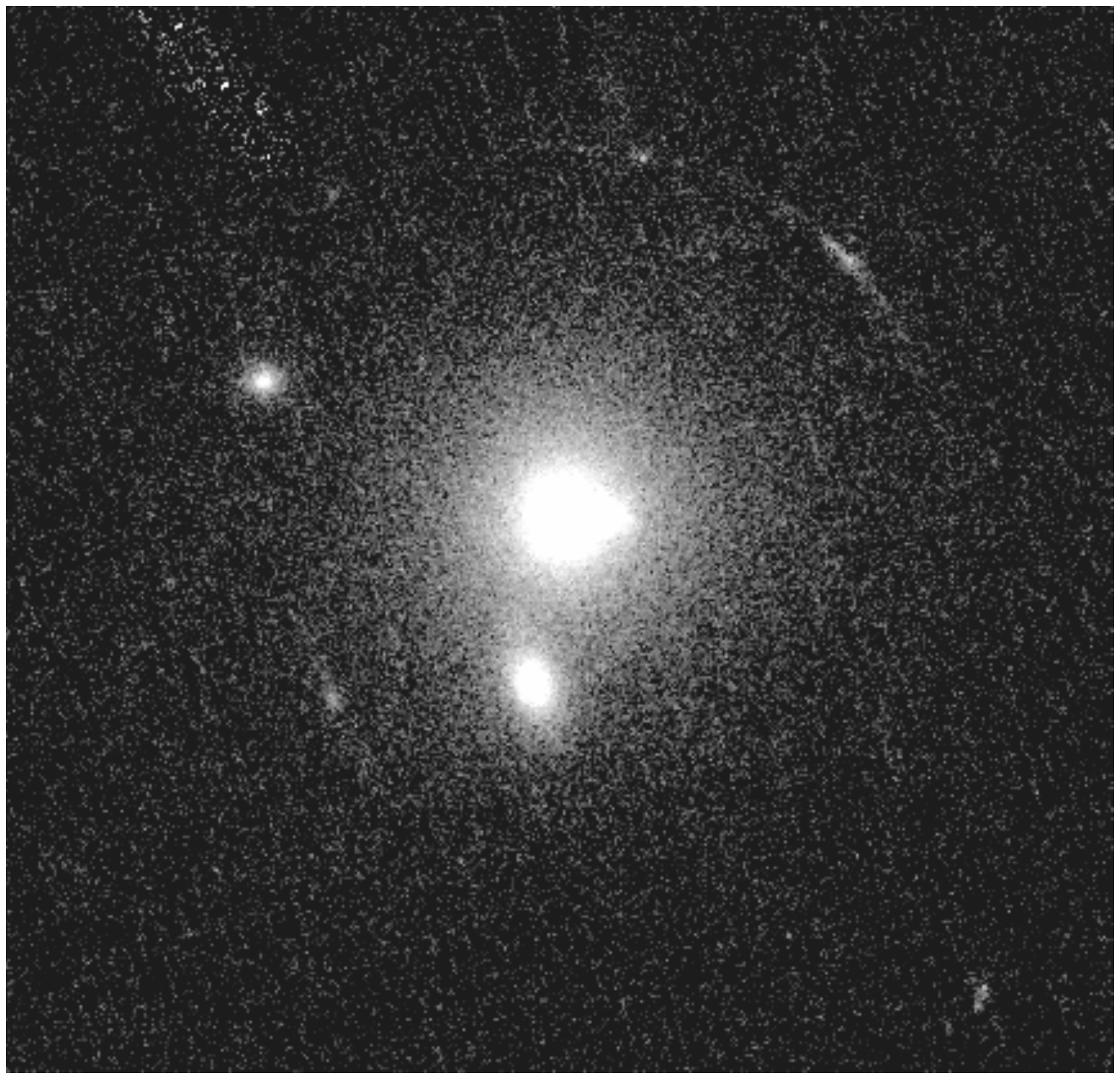}
   \includegraphics[trim=2.3cm 5.1cm 2.3cm 3.5 cm, clip, width=0.30\linewidth]{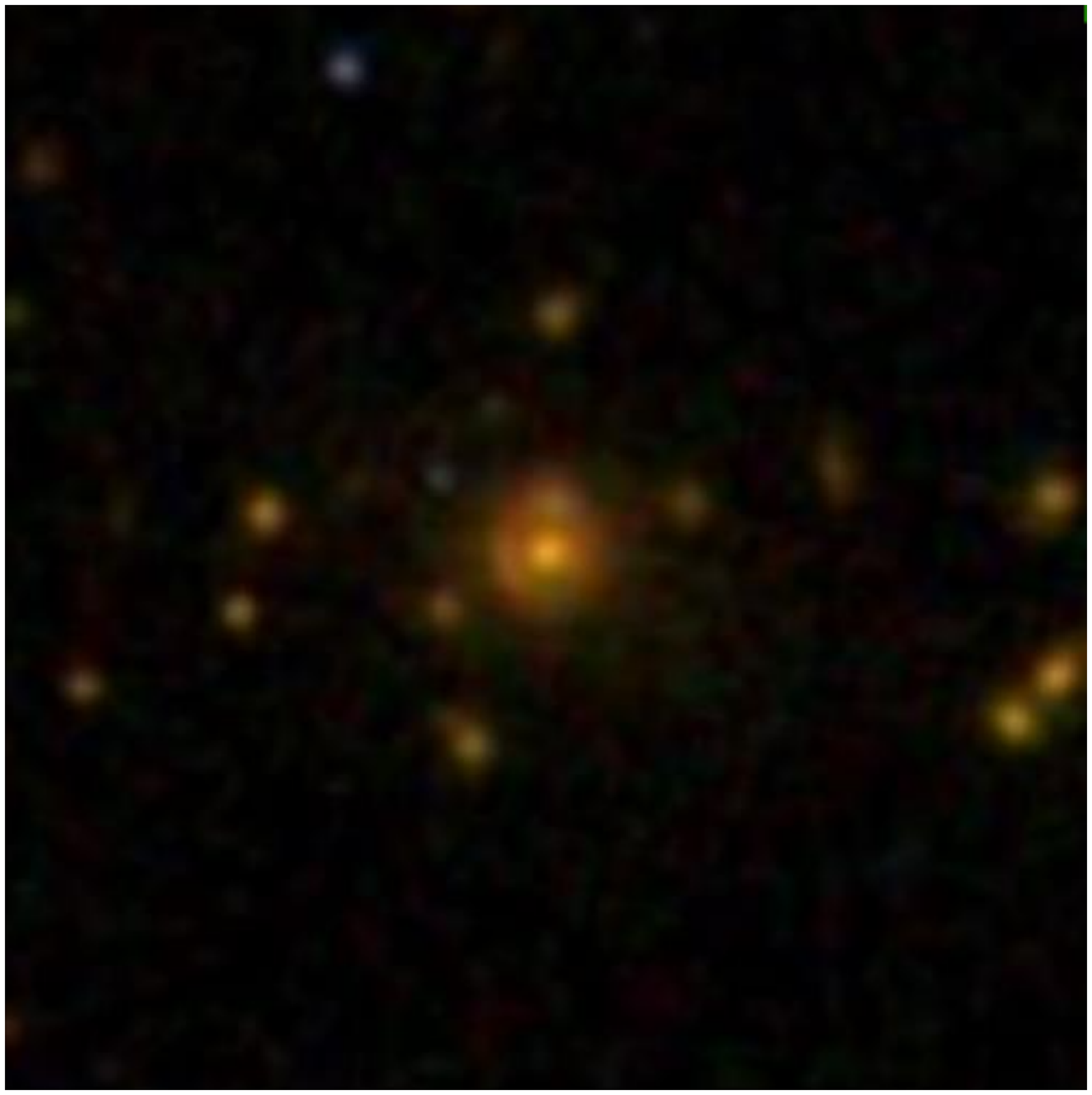}
   \figcaption{Left: HST WFPC2 F814W image of ex-super spiral OGC 0302, showing that it is a disturbed, non-spiral quasar host. The field of view is 24 x 21 arcsec (88 x 77 kpc). 
                    Center: Elliptical galaxy and newly identified gravitational lens OGC 0200 (2MASX J08355126+3926220) imaged by HST ACS/HRC in the F775W filter (Proposal ID 10199).  
                    We initially classified this as a super spiral, mistaking the lens arcs for spiral arms. The field of view is 14 x 14 arcsec (56 x 56 kpc).
                    Right: Newly identified gravitational lens candidate OGC 1565 (2MASX J21531028+1154551).  The field of view is 50 x 50 arcsec (217 x 217 kpc).
   \label{A4}}
\end{figure}

\acknowledgements
We thank Katey Alatalo for insightful discussions regarding the properties of post-merger and post-starburst galaxies.
This work was made possible by the NASA/IPAC Extragalactic Database and the NASA/ IPAC Infrared Science Archive, which are both operated by the Jet Propulsion Laboratory, California Institute of Technology, under contract with the National Aeronautics and Space Administration.  We thank Ben Chan, Marion Schmitz, and the rest of the NED team for useful discussions and
their support of this work.  This publication makes use of data from the {\it Galaxy Evolution Explorer} and the {\it Hubble Space Telescope}, retrieved from the Mikulski Archive for Space Telescopes (MAST). STScI is operated by the Association of Universities for Research in Astronomy, Inc., under NASA contract NAS5-26555. Support for MAST for non-HST data is provided by the NASA Office of Space Science via grant NNX09AF08G and by other grants and contracts. Funding for the Sloan Digital Sky Survey IV has been provided by the Alfred P. Sloan Foundation, the U.S. Department of Energy Office of Science, and the Participating Institutions. SDSS acknowledges support and resources from the Center for High-Performance Computing at the University of Utah. The SDSS web site is www.sdss.org.  SDSS is managed by the Astrophysical Research Consortium for the Participating Institutions of the SDSS Collaboration.
This publication makes use of data products from the Two Micron All Sky Survey, which is a joint project of the University of Massachusetts and the Infrared Processing and Analysis Center/California Institute of Technology, funded by the National Aeronautics and Space Administration and the National Science Foundation. This publication makes use of data products from the {\it Wide-field Infrared Survey Explorer}, which is a joint project of the University of California, Los Angeles, and the Jet Propulsion Laboratory/California Institute of Technology, funded by the National Aeronautics and Space Administration.  

\eject

\clearpage
\LongTables

\begin{deluxetable}{clrccll}
\tablecaption{The Ogle et al. Galaxy Catalog (OGC)}
\tablehead{
\colhead{OGC} & \colhead{NED Name} & \colhead{$L_r$\tablenotemark{a}} & \colhead{r (mag) \tablenotemark{b}} & \colhead{z (NED)\tablenotemark{c}} & \colhead{Morph.\tablenotemark{d}} & \colhead{Spec.\tablenotemark{e}}}
0021  & 2MASX J12220526+4518109       &  19.8 &  16.26 & 0.264357 & E & \\
0022  & B3 1715+425                                 &  19.5 &  15.34 & 0.182900 & E & H$\alpha$, RG  \\
0023  & SDSS J215541.97+123128.5        &  19.1 &  15.68 & 0.193000 & E &  \\  
\nodata  &                                                    &          &            &                 &  & \\
0078  & 2MASX J02295551+0104361        &  13.3 &  16.53 & 0.245472 & E & H$\alpha$ \\
\nodata  &                                                    &          &            &                 &  & \\
1606  & SDSS J121644.34+122450.5         &  8.0 &  17.22 & 0.257144 & SS & Sey1 \\
1607  & SDSS J130454.84+100011.6         &  8.0 &  17.19 & 0.255875 & E &  H$\alpha$ \\
1608  & SDSS J040422.91-054134.9         &  8.0 &  17.27 & 0.250635 & SS & H$\alpha$\\
1609  & 2MASX J12071504+1713488        &  8.0 &  16.87 & 0.221788 & E & \\
1610  & 2MASX J08302311+4744471        &  8.0 &  17.10 & 0.243735 & E & \\
1611  & 2MASX J00380781-0109365         &  8.0 &  16.65 & 0.208695 & S0/Sa & H$\alpha$\\
\hline
1662\tablenotemark{f}  & SDSS J085123.17-002148.7         &  7.9 &  17.59 & 0.295481 & Pec & H$\alpha$  
\enddata
\tablecomments{Table A1 is published in its entirety in the machine-readable format. A portion is shown here for guidance regarding its form and content.}
\tablenotetext{a}{SDSS $r$-band luminosity, k-corrected, and divided by $L^*=5.41 \times 10^{43}$ erg s$^{-1}$.}
\tablenotetext{b}{SDSS $r$-band magnitude.}
\tablenotetext{c}{NED preferred redshift, primarily from SDSS DR6.}
\tablenotetext{d}{Galaxy morphology: E $=$ elliptical,  S0/Sa $=$ super lenticular, SS $=$ super spiral, Pec $=$ peculiar}
\tablenotetext{e}{Galaxy spectrum: H$\alpha$ = H$\alpha$ emission line, Sey1 $=$ Seyfert 1 AGN, RG $=$ radio galaxy}
\tablenotetext{f}{The starburst galaxy OGC 1662 falls below the $8.0L^*$ cutoff of the sample (horizontal line), but we include it in this table, Table 5, and Fig. A3 to point out it's unusual properties.} 
\end{deluxetable}

\begin{deluxetable}{clll}
\tablecaption{Luminous Galaxy Rejects -- Inaccurate Photometry}
\tablehead{
\colhead{OGC} & \colhead{NED Name} & \colhead{Notes}  }

\startdata
0001 & SDSS J101736.97+305101.6 & glare \\ 
0003 & SDSS J091428.98+184250.9 & glare \\ 
0004 & SDSS J102825.27+154757.2 & glare \\
\nodata &                                            & \\
1418 & 2MASX J14352497-0105084 & bright star \\
1425 & 2MASX J10240215+2534587 & glare \\
1585 & 2MASX J09095768+5731219 & bright star \\  
\enddata
\tablecomments{Table A2 is published in its entirety in the machine-readable format. A portion is shown here for guidance regarding its form and content.}
\end{deluxetable}
 
 \begin{deluxetable}{llllll}
\tablecaption{Luminous Galaxy Rejects -- Incorrect Redshifts}
\tablehead{
\colhead{OGC} & \colhead{NED Name} & \colhead{$z$ (NED)} & Ref. (1) & \colhead{ adopted $z$} & Ref. (2)  }

\startdata
0005 & 2MASX J14342221+0706510 & 0.254000 & RK12 & \nodata & \\
0006 & CGCG 045-047                       & 0.192300 & HT10 & \nodata & \\
0008 & 2MASX J13041968+4214150 & 0.201700 & HT10 & 0.03611 & DR12 \\
0011 & UGC 08569                             & 0.140657  & FK99 & 0.02303 & DR12 \\
0012 & 2MASX J16521103+2344396 & 0.178900 & HT10 & 0.03471 & DR12 \\
0013 & 2MASX J11064145+3435219 & 0.158800 & HT10 & \nodata & \\
0014 & LCSB S2225P                        & 0.222100 & RR05 & \nodata & \\
0015 & MCG +08-24-006                   & 0.149600 & HT10 & \nodata & \\
0019 & SBS 0957+569                       & 0.147600 & HT10 & 0.01368 & DR12 \\
0020 & SDSS J082143.27+011423.2 & 0.295711 & GS09 & \nodata & \\ 
0024 & CGCG 208-020                      & 0.132400 & HT10 & 0.02457 & DR12 \\
0026 & 2MASX J15032011+0856496 & 0.168000 & LS10 & \nodata & \\
0030 & 2MASX J01132663+1520072 & 0.203174 & CB09 & 0.04232 & DR12 \\
0036 & 2MASX J01122497+1538371 & 0.217106 & CB09 & 0.04423 & DR12 \\
0038 & 2MASX J14153929+2313477 & 0.252000 & BG06 & 0.06369 & DR12\\
0045 & 2MASX J16064936+3202464 & 0.135400 & HT10 & 0.11550 & DR12 \\ 
0052 & VCC 0838                               & 0.295417 & CG01 & \nodata & \\
0056 & KUG 1320+255                       & 0.143843 & RG01 & 0.03322 & DR12 \\
0069 & 2MASX J09075107+4401533 & 0.114500 & HT10 & 0.04878 & DR12 \\
0102 & SDSS J135128.05+091559.7 & 0.294330 & DR6   & 0.06062 & DR12 \\
0172 & 2MASX J21472854-0738031 & 0.148299 & 6dF    & 0.05971 & DR12 \\
1035 & SDSS J141655.53+234018.2 & 0.281682 & DR13 & 0.11548 & DR12 \\  
1274 & UGC 10782                             & 0.086000 & VC04 & 0.036     & VC04 \\ 
1343 & 2MASX J12110100+3048346 & 0.192703 & DR6   & 0.12861 & DR12 
\enddata
\tablenotetext{a}{DR6,12,13 $=$ SDSS Data Release 6,12, 13; 6dF $=$ 6dF Galaxy Survey; BG06 $=$ \cite{bg06}; CB09 $=$ \cite{cb09}; CG01 $=$ \cite{cgw01}; FK99 $=$ \cite{fk99};  GS09 $=$ \cite{gs09}; 
                            HT10 $=$ \cite{ht10}; LS10 $=$ \cite{ls10}; RG01 $=$ \cite{rg01}; RK12 $=$ \cite{rk12};  RR05 $=$ \cite{rr05};; VC04 $=$ \cite{vc04}}
\end{deluxetable}

\begin{deluxetable}{cllll}
\tablecaption{Luminous Galaxy Rejects -- Overlapping Objects}
\tablehead{
\colhead{OGC} & \colhead{Object 1 (NED) } & \colhead{Type 1} & \colhead{Object2}  & \colhead{Type 2}
}
\startdata
0002 & SDSS J104819.41+123745.8  & compact  & NGC 3384 & E \\
0010  & SDSS J104248.72+132710.7 & compact  & UGC 05832 & Irr \\
0025  & SDSS J141309.29+083707.2 & compact  & VV 299a      & Irr \\
0031  & 2MASX J12494045+2546186 & compact  & KUG 1247+260 & dE \\
0048  & 2MASX J07543650+3905307 & dE?         & no name  & compact \\ 
0059  & 2MASX J08540011+5751327 & E             &no name &S, edge-on  \\ 
0146 & SDSS J092209.32+335057.4  & compact  & UGC 04974 & E \\
0191 & VIII Zw 125 NOTES02             & QSO        & SDSS J110717.65+080435.3 & E \\
0379 & KUG 0907+332                        & E             & SDSS J091031.83+330327.9 & Irr \\
0706 & SDSS J142418.44+293238.4  & E             & SDSS J142418.71+293232.9 & S \\ 
1176 & SDSS J144700.16+011504.9  & F9 star (cz=9 km/s) &no name  & G \\    	 
1322 & SDSS J014503.26-001800.0  & S              & UGC 01225 NED02 & Irr \\ 
1360 & SDSS J123433.28+181154.8 & E              & NGC 4539 & S0 \\   
1406 & 2MASX J11523751+1527539 & E              & SDSS J115237.63+152759.1 & S \\ 
1497  & SDSS J020707.48-082726.0 & compact   & GALEXASC J020707.37-082723.5 & Irr\\
1565 & 2MASX J21531028+1154551  & GLens candidate & no name & lensed galaxy 
\enddata
\tablenotetext{}{Galaxy morphology:   compact $=$ star or compact galaxy, G $=$ unclassified galaxy,  S $=$ spiral, S0 $=$ lenticular, E $=$ elliptical, dE $=$ dwarf elliptical, Irr $=$ irregular galaxy, QSO $=$ quasi-stellar object, GLens $=$ gravitational lens.}
\end{deluxetable}

\eject

\begin{deluxetable}{cccrrccll}
\tablecaption{OGC Super Spirals, Lenticulars, and Post-mergers}
\tablehead{
\colhead{OGC} & \colhead{NED Name} & \colhead{$z$(SDSS)\tablenotemark{a}} & \colhead{$L_r$\tablenotemark{b}} & \colhead{$D$\tablenotemark{c}} & \colhead{$\log M_\mathrm{stars}$\tablenotemark{d}} & \colhead{$\log$ SFR\tablenotemark{e} } & \colhead{Morph.\tablenotemark{f}} & \colhead{Spect.\tablenotemark{g} }
}
\startdata

0065 & 2MASX J10301576-0106068 & 0.28228 & 13.9 &  81 & 11.65 & 1.21 &  bar     & \\ 
0139 & 2MASX J16394598+4609058 & 0.24713 & 12.0 &134 & 11.71 & 1.31 &            & H$\alpha$ \\ 
0170 & 2MASX J10100707+3253295 & 0.28990 & 11.6 &  87 & 11.65 & 1.08 & bar      & H$\alpha$ \\ 
0179 & SDSS J213701.13-064447.0  & 0.29065 & 11.5 &  76 & 11.88 & 0.34 &             &  \\ 
0217 & 2MASX J13275756+3345291 & 0.24892 & 11.2 &  69 & 11.52 & 1.50 & bar      & H$\alpha$, sb \\ 
0222 & 2MASX J12220815+4844557 & 0.29861 & 11.1 &  95 & 11.78 & 0.44 &            & \\ 
0256 & 2MASX J11593546+1257080 & 0.26353 & 10.9 &  87 & 11.56 & 0.90 &            & H$\alpha$ \\
0290 & 2MASX J12343099+5156295 & 0.29592 & 10.6 &  62 & 11.55 & 1.26 &            & Sey1 \\  
0293 & 2MASX J13044128+6635345 & 0.28862 & 10.6 &  77 & 11.74 & 0.69 &            & H$\alpha$ \\ 
0299 & 2MASX J09094480+2226078 & 0.28539 & 10.5 &  83 & 11.84 & 0.65 &            &    \\ 
0306 & SDSS J122100.48+482729.1 & 0.29966 & 10.5 &  75 & 11.72 & 0.79 &            &    \\  
0345 & 2MASX J09260805+2405242 & 0.22239 & 10.3 &  81 & 11.72 & 1.17 &            & H$\alpha$ \\ 
0388 & 2MASX J17340613+6029190 & 0.27596 & 10.1 &  64 & 11.57 & 0.84 &            &    \\ 
0441 & SDSS J095727.02+083501.7 & 0.25652 &   9.9 &  88 & \nodata& 0.81 &            &    \\ 
0454 & 2MASXi J1003568+382901 & 0.25860    &   9.9 &  56 & 11.68 & 1.28 &            & H$\alpha$, sb \\ 
0516 & 2MASX J14475296+1447030 & 0.22069 &   9.7 &  95 & 11.65 & 0.89 &            & LINER\\ 
0543 & 2MASX J09470010+2540462 & 0.10904 &   9.6 &  99 & 11.65 & 1.05 & bar      & Sey1\\ 
0574 & SDSS J121148.70+662514.4 & 0.23789 &   9.5 &  63 & 11.75 & 0.96 &             & H$\alpha$  \\ 
0586 & 2MASX J11535621+4923562 & 0.16673 &   9.5 &  90 & 11.63 & 1.27 &             & Sey2 \\ 
0595 & 2MASX J07550424+1353261 & 0.22264 &   9.5 &  77 & 11.53 & 0.96 & bar       & \\ 
0612 & SDSS J093540.34+565323.8 & 0.29636 &   9.4 &  92 & 11.98 & \nodata & bar       &  \\ 
0623 & 2MASX J09011007+2454570 & 0.25232 &   9.4 &  74 & 11.55 & 1.03 &             & Sey1\\ 
0637 & 2MASX J15575566+4322473 & 0.20641 &   9.3 &  80 & 11.60 & 0.72 &             & H$\alpha$\\ 
0696 & SDSS J102154.85+072415.5 & 0.29061 &   9.2 &  70 &\nodata& 0.90 &             & H$\alpha$ \\ 
0704 & 2MASX J03460305+0100064 & 0.18605 &   9.2 &  83 & 11.46 & 0.91 &            &  \\ 
0713 & 2MASX J08265512+1811476 & 0.26545 &   9.2 &  82 & 11.64 & 0.93 & bar      & \\ 
0749 & 2MASX J15591044+3826290 & 0.29073 &   9.1 &  70 & 11.56 & 0.98 & bar      &  Sey1 \\ 
0753 & 2MASX J08022926+2325161 & 0.27152 &   9.1 &  66 & 11.35 & 0.98 &            & H$\alpha$\\ 
0755 & SDSS J113800.88+521303.9 & 0.29593 &    9.1 &  64 & 11.59 & 0.79 &            &  \\ 
0789 & 2MASX J08542169+0449308 & 0.15679 &   9.0 &  86 & 11.58 & 1.13 & bar      & H$\alpha$ \\ 
0799 & 2MASX J10472505+2309174 & 0.18256 &   9.0 &  72 & 11.65 & 1.01 & bar      & \\ 
0800 & 2MASX J11191739+1419465 & 0.14377 &   9.0 &  71 & 11.55 & 0.98 &             &      \\ 
0804 & SDSS J135546.07+025455.8 & 0.23884 &   9.0 &  84 &\nodata& 0.77 &             &      \\ 
0830 & SDSS J141754.96+270434.4 & 0.15753 &   9.0 &  69 & 11.69 & 0.90 &             &     \\ 
0909 & 2MASX J14381016+5030122 & 0.24665 &   8.8 &  66 & 11.53 & 0.32 &             &      \\ 
0926 & 2MASX J10304263+0418219 & 0.16092 &   8.8 &  70 & 11.66 & 1.36 &             & H$\alpha$\\ 
0928 & 2MASX J12374668+4812273 & 0.27245 &   8.8 &  66 & 11.46 & 1.08 &             & H$\alpha$ \\ 
0968 & 2MASX J09312816+4424163 & 0.21940 &   8.7 &  65 & 11.50 & 1.29 & bar       & Sey2, sb \\ 
0975 & 2MASX J11410001+3848078 & 0.26770 &   8.7 &  72 & 11.54 & 1.08 &             &  \\ 
0983 & SDSS J153618.97+452246.8 & 0.23618 &   8.7 &  80 & 11.43 & 0.71 &             &  \\ 
0984 & SDSS J133737.88+494015.6 & 0.27233 &   8.7 &  73 & 11.74 & 0.35 &             & H$\alpha$\\ 
0995 & 2MASX J14440406+2029072 & 0.24820 &   8.7 &  76 & 11.83 & 0.31 &             & H$\alpha$ \\
1023 & 2MASX J09254889+0745051 & 0.17227 &   8.6 &  68 & 11.62 & 0.85 &             & H$\alpha$\\ 
1046 & 2MASX J09362208+3906291 & 0.28293 &   8.6 &  70 & 11.42 & 0.85 &             &  \\ 
1088 & SDSS J140138.37+263527.6 & 0.28396 &   8.5 &  78 &\nodata& 0.90 &             & H$\alpha$\\ 
1094 & SDSS J163357.99+172839.5 & 0.26691 &   8.5 &  77 &\nodata & 1.08 &             & H$\alpha$ \\ 
1107 & 2MASX J12072497-0150416 & 0.20957 &   8.5 &  69 & 11.46 & 0.91 & bar       &   \\ 
1166 & 2MASX J22295446-0921345 & 0.27954 &   8.4 &  57 & 11.49 & 0.46 &             &  \\
1182 & 2MASX J00495939-0853413 & 0.12181 &   8.4 &  66 & 11.84 & 0.82 &             & H$\alpha$\\ 
1196 & SDSS J154950.91+234444.1 & 0.26208 &   8.4 &  69 &\nodata& 0.99 &             &  H$\alpha$\\ 
1250 & 2MASX J12321515+1021195 & 0.16588 &  8.3 &  71 & 11.42 & 0.88 &             &     \\ 
1268 & 2MASX J12005393+4800076 & 0.27841 &  8.3 &  63 & 11.46 & 1.04 &             & H$\alpha$ \\ 
1273 & 2MASX J07380615+2823592 & 0.23091 &  8.3 &  77 & 11.46 & 0.97 & bar      &   \\
1287 & 2MASX J07404205+4332412 & 0.17828 &   8.3 &  69 & 11.62 & 0.90 &            & H$\alpha$ \\ 
1304 & 2MASX J16014061+2718161 & 0.16440 &   8.3 &  82 & 11.80 & 1.33 &           & H$\alpha$  \\ 
1312 & SDSS J143447.86+020228.6 & 0.27991 &   8.2 &  75 & 11.68 & 1.26 &             & H$\alpha$, sb \\ 
1323 & SDSS J112928.74+025549.9 & 0.23960 &   8.2 &  70 & 11.70 & 0.94 & bar       &   \\ 
1329 & 2MASX J16273931+3002239 & 0.25990 &   8.2 &  86 & 11.75 & 0.48 &             & H$\alpha$\\
1337 & SDSS J093921.25+260709.8 & 0.27487 &   8.2 &  56 & 11.70 & 0.68 &             & H$\alpha$ \\
1352 & SDSS J101603.97+303747.9 & 0.25191 &   8.2 &  69 & 12.00 & \nodata&             & H$\alpha$ \\ 
1375 & 2MASX J00155012-1002427 & 0.17601 &   8.2 &  68 & 11.47  & 0.75 &              & H$\alpha$ \\ 
1379 & 2MASX J09373465+1036552 & 0.17946 &   8.2 &  90 & 11.58 & 1.04 & bar       & H$\alpha$ \\ 
1395 & 2MASX J13103930+2235023 & 0.23123 &   8.1 &  66 & 11.53 & 0.86 &             &      \\ 
1409 & SDSS J151721.02+603302.6 & 0.28232 &   8.1 &  70 & 11.62 & 0.79 &             & H$\alpha$ \\ 
1420 & 2MASX J13475962+3227100 & 0.22306 &   8.1 &  88 & 11.50 & 0.93 &             &       \\ 
1423 & SDSS J215250.41+122159.2 & 0.27310 &   8.1 &  61 & \nodata & 0.90 &             & H$\alpha$ \\ 
1428 & 2MASX J11162790+3813476 & 0.23350 &   8.1 &  77 & 11.34 & 0.89 &             &      \\ 
1450 & SDSS J132743.82-031323.1 & 0.29502 &   8.1 &   64 &\nodata & 0.91 &             & H$\alpha$  \\
1457 & 2MASX J09381666+1044508 & 0.23897 &  8.1 &  72 & 11.37 & 1.26 &              & H$\alpha$, sb \\ 
1464 & 2MASX J10041606+2958441 & 0.29844 &  8.1 &  57 & 11.47 & 1.39 &              & H$\alpha$, sb \\ 
1500 & 2MASX J10095635+2611324 & 0.24089 &  8.1 &  64 & 11.44 & 0.96 &              & H$\alpha$ \\ 
1501 & 2MASX J09334777+2114362 & 0.17219 &  8.1 &  64 & 11.62 & 1.43 &              & QSO, sb \\ 
1512 & SDSS J122944.64+272306.3 & 0.27573 &  8.0 & 101 & 11.48 & 0.89 &             & H$\alpha$, sb \\ 
1514 & SDSS J080317.08+325932.6 & 0.24848 &  8.0 &   55 & 11.89 &  0.24 &              &      \\ 
1520 & 2MASX J12354859+3919078 & 0.23706 &  8.0 &   63 & 11.64 & 1.46 &             & H$\alpha$, sb \\ 
1544 & 2MASX J14472834+5908314 & 0.24551 &  8.0 &   68 & 11.58 & 0.80 & bar      & H$\alpha$ \\ 
1546 & 2MASX J13435549+2440484 & 0.13725 &  8.0 &   66 & 11.59 & 0.66 &            & H$\alpha$ \\ 
1549 & 2MASX J08464747+0446053 & 0.24145 &  8.0 &   76 & 12.01 & \nodata &             &     \\ 
1554 & 2MASX J13422833+1157345 & 0.27873 &  8.0 &   57 & 11.53 & 0.94 &             & H$\alpha$ \\ 
1559 & CGCG 122-067                       & 0.08902 &  8.0 &   81 & 11.71 & 0.93 &             & H$\alpha$ \\ 
1562 & SDSS J163202.04+464545.7 & 0.29491 &  8.0 &   67 & 11.64 & 0.61 &              & H$\alpha$\\ 
1600 & SDSS J115155.92+104634.7 & 0.28305 &  8.0 &   67 & 11.63 & 0.22 &              & H$\alpha$  \\ 
1606 & SDSS J121644.34+122450.5 & 0.25694 &  8.0 &   78 & \nodata & 0.95 & bar      & Sey1\\ 
1608 & SDSS J040422.91-054134.9 & 0.25055 &  8.0 &   80 & 11.51 & 0.78  &              & H$\alpha$ \\ 
\hline
0044 & 2MASX J14072225+1352512 & 0.29372 & 15.1 & 85 & 11.92 & 0.47 &  S0/Sa & \\ 
0073 & 2MASX J10405643-0103584 & 0.25024 & 13.4 &  82 & 11.81 & 0.49 &  S0/Sa & \\ 
0265 & SDSS J115052.98+460448.1 & 0.28946 & 10.8 &  88 & 11.80 & 0.10 &  S0/Sa &  \\ 
0280 & 2MASX J09572689+4918571 & 0.24144 & 10.7 &106 & 11.81 & 0.60 & S0/Sa &  \\ 
0425 & 2MASX J21160443-0702228  & 0.19082 & 10.0 &  77 & 11.85 & 0.60 & S0/Sa & H$\alpha$ \\
0581 & 2MASX J13423113+0021440 & 0.24342 &   9.5 &142 & 11.80 & 0.64 & S0/Sa   &      \\
1002 & 2MASX J10535662+5909155 & 0.19896 &   8.7 &  84 & 11.69 & 0.76 & S0/Sa  & H$\alpha$  \\
1220 & 2MASX J08164326+4702216 & 0.29529 &  8.3 &  62 & 11.55 & 0.63 & S0/Sa & H$\alpha$ \\
1270 & SDSS J125157.99+305422.3 & 0.23065 &  8.3 &  78 & 11.77 & 0.48 & S0/Sa & H$\alpha$ \\
1291 & SDSS J090317.22-000758.9 & 0.29726 &   8.3 &   72 & 11.77 & 0.33 & S0/Sa  & H$\alpha$ \\
1381 & 2MASX J08093749+2316385 & 0.27291 &   8.2 &  81 & 11.65 & 0.33 & S0/Sa  &       \\ 
1386 & 2MASX J13382172+0929423 & 0.24302 &   8.2 &  92 & 11.59 & 0.68 & S0/Sa  &  \\ 
1526 & 2MASX J11414166+0223211 & 0.23354 &  8.0 &   89 & 11.58 & 0.39 & S0/Sa  & H$\alpha$ \\ 
1535 & 2MASX J11160517+3303477 & 0.20616 &  8.0 &   94 & 11.65 & 0.36 & S0/Sa  &   \\ 
1611 & 2MASX J00380781-0109365 & 0.20828 &  8.0 &   84 & 11.70 & 0.74  & S0/Sa    & H$\alpha$  \\ 

\hline
0247 & SDSS J081953.52+041409.2  & 0.296625 & 10.9 & 43 & 11.56 & 0.90  & Pec             & K+A + Sey1 \\ 
0331 & SDSS J091318.25+492556.3  & 0.296414 & 10.4 & 43 & 11.52 & 0.80 & Pec             & K+A + Sey1 \\   
0624 & 2MASX J13245634+6219585  & 0.237397 &  9.4 & 69 & 11.64 & 1.06  & Pec             & K+A ? + Sey1 \\
0707 & 2MASX J11304267+1538467  & 0.298136 &  9.2 & 61 & 11.50 &  0.60 & Pec             & K+A \\
0783 & SDSS J102629.10+094519.7  & 0.262280 &  9.0 & 72 & 11.69 & 0.82 & Pec             & K+A \\
0892 & SDSS J095543.25+111715.9   & 0.299029 &  8.8 & 66 & 11.70 & 0.18 & Pec             & H$\alpha$ \\
0902 & 2MASX J23591456+1351308  & 0.247131 &  8.8 & 55 & 11.56 & 1.34 & Pec             & H$\alpha$, sb  \\     
0973 & 2MASX J13412783+2851280  & 0.294948 &  8.7 & 47 & 11.59 & 1.03 & Pec             & K+A+H$\alpha$ \\
1056 & SDSS J120050.60-012755.6   & 0.267160 &  8.6 & 117 & \nodata & 0.57 & Pec             & \\         
1058 & 2MASX J12383963+6413430  & 0.265032 &  8.6 & 82 & 11.57 & 1.21 & Pec             & H$\alpha$, sb \\    
1174 & 2MASX J11310763+0224271  & 0.257493 &  8.4 & 65 & 11.50 & 0.88 & Pec             & H$\alpha$ \\  
1377 & SDSS J134719.23+114915.1  & 0.279850 &  8.2 & 66 & 11.46 & 0.77 & Pec              & K+A + Sey2 \\
1413 & MCG +09-25-047                     & 0.244470 &  8.1 & 67 & 11.85 & $<1.90$ & Pec             & K+A + Sey2 \\
1490 & 2MASX J08164043+3340182  & 0.238297 &  8.1 & 58 & 11.34 & 0.88 & Pec             & K+A + [N II] \\     
1662 & SDSS J085123.17-002148.7   & 0.295481 &  7.9 & 72 & \nodata & 1.19 & Pec              & H$\alpha$, sb 
\enddata

\tablenotetext{a}{SDSS DR9 redshift.}
\tablenotetext{b}{$L/L^*$ (Sloan $r$-band).}
\tablenotetext{c}{Isophotal diameter (kpc) at $r = 25.0$ mag arcsec$^{-2}$.}
\tablenotetext{d}{log of mass in stars ($M_\odot$)}
\tablenotetext{e}{log of star formation rate ($M_\odot$ yr$^{-1}$) }
\tablenotetext{f}{Morphology. Lenticular galaxies are denoted S0/Sa. Galaxies with stellar bars are indicated as such. Horizontal lines in the table separate super spirals, super lenticulars, and super post-mergers.}
\tablenotetext{g}{Notes on SDSS spectroscopy. H$\alpha$ indicates detection of that line in the SDSS spectrum. AGNs are marked as Seyfert 1 (Sey1), Seyfert 2 (Sey2), LINER, or QSO.  Galaxies with SSFR $ >1/13.7$ Gyr are marked sb. }

\end{deluxetable}

\begin{deluxetable}{lllll}
\tablecaption{OGC AGN-dominated}
\tablehead{
\colhead{OGC} & \colhead{NED Name} & \colhead{Other Name} & \colhead{Image} &\colhead{Spectrum} }

\startdata
0239 & 2MASX J11552373+1507564  & & AGN host    & QSO \\
0302 & 2MASX J15430777+1937522 &  &  AGN host   & QSO \\ 
0307 & SDSS J143335.34+242039.2  &  & 2E + star?   & \\
0377 & 2MASSi J2342593+134750     &  & AGN host    & QSO \\
0469 & SDSS J150022.77+220027.3  & NVSS J150022+220027 & 2G + star? & \\                   
0615 & 2MASXi J0837247+145819     & ABELL 0689:[REE2012] BCG  & AGN host & BLLac \\          
0646 & 2MASX J21193928+1039326  &  & 2E + 2 stars? & \\      
0889 & 2MASX J08250928+2634381  &  & AGN host     & QSO \\ 
0962 & SDSS J141756.67+254326.2  &  & AGN host     & BL Lac\\
1229 & 2MASX J08574977+0135301  &  & AGN host     & BL Lac \\
1239 & SDSS J145608.63+380038.5  &  & AGN host     & QSO \\
1245 & 2MASX J02354663-0742506   &  & AGN host     & QSO 
\enddata
\end{deluxetable}

\begin{deluxetable}{cllcclllll}
\tablecaption{Super Spiral and Lenticular Cluster and Group Associations}
\tablehead{
\colhead{OGC} &\colhead{NED Name} & \colhead{Redshift} & \colhead{N1\tablenotemark{a}} & \colhead{N10\tablenotemark{b}} & \colhead{Cluster Name} & \colhead{Type}& \colhead{Redshift} & \colhead{ztype\tablenotemark{c}} 
                       &  \colhead{Sep($'$)\tablenotemark{d}}}
\startdata
0044 & 2MASX J14072225+1352512  & 0.293596 & 1 & 23 & GMBCG J211.84274+13.88070      & GClstr   & 0.280850 & PHOT  & 0.000 \\
0073 & 2MASX J10405643-0103584   & 0.250303 & 1 & 13 & SDSS CE J160.241898-01.069106 & GClstr   & 0.254019 & EST    & 0.013 \\
0170 & 2MASX J10100707+3253295  & 0.289913 & 2 & 38 & GMBCG J152.52936+32.89139      & GClstr   & 0.319000 & PHOT & 0.001 \\
0179 & SDSS J213701.13-064447.0   & 0.290697 & 1 &   8 & SDSSCGB 18956                            & GGroup & 0.291000 & SPEC & 0.054 \\
0280 & 2MASX J09572689+4918571  & 0.241492 & 5 & 26 & MaxBCG J149.36205+49.31591     & GClstr   & 0.237650 & PHOT & 0.000  \\
0293 & 2MASX J13044128+6635345  & 0.288630 & 1 &   4 & MaxBCG J196.17181+66.59301     & GClstr   & 0.226850 & PHOT & 0.000 \\
0299 & 2MASX J09094480+2226078  & 0.285386 & 2 & 17 & GMBCG J137.43670+22.43538      & GClstr   & 0.303000 & PHOT & 0.000 \\
0345 & 2MASX J09260805+2405242  & 0.222451 & 1 & 36 & WHL J092608.1+240524                 & GClstr   & 0.178000 & PHOT & 0.000 \\
0388 & 2MASX J17340613+6029190  & 0.275807 & 1 &   2 & SDSSCGB 59704                            & GGroup & 0.276000 & SPEC & 0.450 \\   
0516 & 2MASX J14475296+1447030  & 0.220592 & 1 & 33 & ABELL 1971                                     & GClstr   & 0.208600  & SPEC & 1.298 \\
         &                                                  &                 &    &      & MaxBCG J221.98726+14.75906      & GClstr   & 0.216050 & PHOT & 1.775 \\
         &      				           &                 &    &      & WHL J144756.9+144532                  & GClstr    & 0.203600 & PHOT & 1.778 \\
0581 & 2MASX J13423113+0021440  & 0.243520 & 1 & 47 & SDSSCG 110                                    & GGroup & 0.243400 & SPEC & 0.439 \\    
         &                                                  &                 &    &      &  SDSS CE J205.645691+00.368013 & GClstr   & 0.231327 & EST    & 1.031 \\                            
0586 & 2MASX J11535621+4923562  & 0.166892 & 3 & 70 & OGC 0586 CLUSTER                       & GClstr    & 0.166187 & SPEC & 0.000 \\
0612 & SDSS J093540.34+565323.8  & 0.296393 & 1 & 21 & ZwCl 0932.1+5708                            & GClstr    & \nodata   & \nodata & 1.186 \\ 
0637 & 2MASX J15575566+4322473  & 0.206452 & 1 & 18 & MaxBCG J239.48210+43.37988      & GClstr    & 0.202550 & PHOT  & 0.000 \\ 
0704 & 2MASX J03460305+0100064  & 0.186147 & 1 & 167 & WHL J034603.0+010006                & GClstr    & 0.181200 & PHOT  & 0.008 \\
0755 & SDSS J113800.88+521303.9  & 0.296018 & 1 & 16 & SDSSCGB 65403                              & GGroup & \nodata    & \nodata & 1.104 \\
1002 & 2MASX J10535662+5909155 & 0.198533 & 1 & 26  & WHL J105356.6+590915                   & GClstr   & 0.210650 & PHOT  & 0.001 \\
1023 & 2MASX J09254889+0745051  & 0.172306 & 2 & 22 & GMBCG J141.45380+07.75151       & GClstr   & 0.129650 & PHOT   & 0.000 \\ 
         &                                                  &                 &    &      & MSPM 09586                                    & GClstr   & 0.134320 & SPEC   & 1.323 \\ 
         &				                   &                 &    &      & SDSSCGA 00090                              & GGroup & 0.134000 & SPEC  & 1.872 \\ 
1088 & SDSS J140138.37+263527.6  & 0.284036 & 1 &  2  & ZwCl 1359.5+2650                            & GClstr   & \nodata   &  \nodata  & 1.902 \\ 
1166 & 2MASX J22295446-0921345   & 0.279639 & 1 &  2  & MaxBCG J337.47710-09.35962       & GClstr   & 0.237650 & PHOT   & 0.000 \\  
1268 & 2MASX J12005393+4800076  & 0.278617 & 2 & 47 & GMBCG J180.22479+48.00211       & GClstr   & 0.252000 &  PHOT  & 0.001 \\ 
1270 & SDSS J125157.99+305422.3  & 0.230703 & 3 & 14  & MaxBCG J192.99166+30.90620     & GClstr   & 0.283550 & PHOT   & 0.001 \\
1304 & 2MASX J16014061+2718161  & 0.164554 & 3 & 164 & GMBCG J240.41924+27.30444     & GClstr   & 0.193000 & PHOT   & 0.000 \\
         &				                   &                 &    &       & MaxBCG J240.43568+27.30263      & GClstr   & 0.164750 & PHOT  & 0.883 \\
         &				                   &                 &    &       & WHL J160144.6+271809                  & GClstr   & 0.162200 & PHOT  & 0.892 \\ 				   
1329 & 2MASX J16273931+3002239 & 0.259761 &  1 & 19 & GMBCG J246.91981+30.01418       & GClstr   & 0.261000 & PHOT  & 1.588 \\  
1420 & 2MASX J13475962+3227100  & 0.223113 & 1 & 16 & SDSSCGB 16827                              & GGroup & \nodata   &  \nodata  & 0.748 \\ 
1520 & 2MASX J12354859+3919078  & 0.237013  & 1 & 13 &  SDSSCGB 36014                            & GGroup & \nodata   & \nodata  & 1.743 \\  
 1549 & 2MASX J08464747+0446053 & 0.241509  & 1 & 20 & WHL J084647.5+044605                  & GClstr   & 0.237650 & PHOT  & 0.000 \\
1559 & CGCG 122-067                        & 0.089008  & 5 & 310 & MSPM 05544                                  & GClstr   & 0.089190 & SPEC  & 0.001 
\enddata
\tablenotetext{a}{Number of galaxies within 1 Mpc and 500 km s$^{-1}$.}
\tablenotetext{b}{Number of galaxies within 10 Mpc and 5000 km s$^{-1}$.}
\tablenotetext{c}{Redshift type, from NED or reference therein. EST--estimated, PHOT--photometric, and SPEC--spectroscopic.}
\tablenotetext{d}{Separation (in arcminutes) of the cluster or group catalog position in NED from the super spiral or super lenticular.  In many cases the separation is zero because
                           the brightest galaxy (i.e. the super spiral) position was apparently used to define the cluster position.}
\end{deluxetable}

\eject

\begin{deluxetable}{cccccrr}
\tablecaption{Super Spiral and Lenticular Mergers and Interacting Pairs}
\tablehead{
\colhead{OGC} & \colhead{Merger}  & \colhead{Features} & \colhead{Companion}  & $z_2$ & \colhead{Sep. ($\arcsec$)} &\colhead{Sep. (kpc) \tablenotemark{a}}}
\startdata
0044 & 2 nuclei       &              & SDSS J140722.29+135250.3 & 0.28601 & 2. & 9. \\  
0290 & major pair    &  $\nu$ shape  & SDSS J123429.74+515639.8 &                &16. & 72.\\
0299 & major triple  &             & SDSS J090944.05+222605.4 &                & 11. & 46.\\
         &                     &             & SDSS J090944.08+222632.2 & 0.28380  & 27. & 115.\\
0425 & 2 nuclei       & debris  &                                                 &                & 2.  & 6.\\ 
0516 & major triple  &             & SDSS J144753.90+144701.2 &                & 14. & 48.\\
         &                     &             & SDSS J144752.76+144728.7 &                & 27. & 95.\\
0574 & 2 nuclei       &             & 2MASX J12114871+6625146 &                & 3. & 12. \\
0581 & major pair    &             & SDSS J134229.89+002138.1 &                &19. & 74. \\ 
0612 & 2 nuclei        &             & SDSS J093540.00+565324.0 &               & 3. & 13. \\
0749 & 2 nuclei        &             & SDSS J155910.19+382626.9 &                & 4. & 15. \\
0755 & 2 nuclei        & ring      &                                                &                & 1. & 4. \\
0789 & 2 nuclei        & tail       &                                                &                & 5.  & 14. \\
0799 & major pair    & ring      & SDSS J104723.75+230923.4 &                &19. & 58. \\
0909 & 2 nuclei        &             &                                                &                & 4. & 15. \\
0968 & 2 nuclei        &             &                                                &                & 3. & 11. \\
0983 & 2 nuclei        & ring      & SDSS J153618.68+452238.8 &                & 9. & 32. \\
0984 & major pair    &             & SDSS J133740.61+494023.5 &  0.27221 & 28. & 115.\\
1002 & 2 nuclei        &             &                                                 &                & 3. & 9. \\ 
1088 & 2 nuclei        & tail       & SDSS J140138.83+263530.6 &                & 7. & 30. \\
1094 & 2 disks         &             & SDSS J163357.74+172836.1 &                & 5. & 20. \\
1182 & 2 nuclei        &             & SDSS J004959.22-085332.0  &                & 10. & 21. \\
1196 & minor pair    & ring      & SDSS J154949.79+234452.2  &                & 17. & 69. \\
1250 & minor pair    &             & SDSS J123212.17+102121.9  &                & 44.  & 125. \\
1304 & 2 nuclei        &            &                                                  &                 &3. & 9. \\
1329 & nest             &             & SDSS J162739.16+300217.7  &                & 7. & 28. \\
         &                     &             & SDSS J162739.98+300227.5  &                & 9. & 38. \\
1352 & minor pair    &             & SDSS J101605.14+303751.3  &                &16. & 61. \\
1375 & 2 nuclei        & tail       & SDSS J001550.69-100243.4   &                & 8. & 25. \\
1395 & 2 nuclei        & tail       & SDSS J131039.16+223506.9  &                & 5. & 18. \\
1409 & 2 nuclei        & ring      & SDSS J151720.48+603301.9  &                & 4. & 17. \\
1420 & 2 nuclei        &             &                                                 &                & 5. & 17. \\
1423 & minor  pair   & $\nu$ shape   & SDSS J215251.02+122156.1  &                & 10. & 40. \\
1457 & major pair    & tail        & SDSS J093815.80+104500.9 &                & 16. & 60. \\
1464 & 2 nuclei        &             & SDSS J100416.17+295839.9 &                & 5. & 21. \\
1500 & 2 nuclei        & ring      &                                                 &                & 4. & 14. \\
1501 & 2 nuclei        & debris  & SDSS J093347.52+211434.0  &                & 4. & 12. \\
1514 & 2 nuclei        &             & SDSS J080316.86+325930.9 &                & 3. & 13. \\
1549 & 2 nuclei        & debris  & SDSS J084647.56+044559.9 &                & 5. & 20. \\
1554 & major pair    & ring      & SDSS J134227.34+115707.0  &                & 31. & 131. \\
1559 & 2 nuclei        & debris  &                                                 &                & 4. & 6. \\
1562 & 2 nuclei        & tail       &                                                 &                & 1. & 4. \\
1600 & 2 nuclei        & tail       &                                                 &                & 2. & 9. \\
1608 & minor pair    &             & SDSS J040423.64-054135.7  &                &11. & 43. 
\enddata
\tablenotetext{a}{Projected separation.}
\end{deluxetable}

\eject


\begin{thebibliography}

\bibitem[Alatalo et al. (2017)]{abl17} Alatalo, K. et al. 2017, ApJ, 843, 9

\bibitem[Alatalo et al. (2014)]{asa14} Alatalo, K., Cales, S. L., Appleton, P. N. et al. 2014, ApJ, 794, L13


\bibitem[Barkhouse et al. (2006)]{bg06} Barkhouse et al. 2005, ApJ, 645, 955

\bibitem[Bamford et al. (2009)]{bnb09} Bamford, S. P. et al. 2009, MNRAS, 393, 1324


\bibitem[Bell \& de Jong (2000)]{bd00} Bell, E. F., \& de Jong, R. S. 2000, MNRAS, 312, 497

\bibitem[Bell et al. (2003)]{bmk03} Bell, E. F., McIntosh, D. H., Katz, N., \& Weinberg, M. D. 2003, ApJS, 149, 289

\bibitem[Blanton et al. (2003)]{bhb03} Blanton, M. R., Hogg, D. W., Bahcall, N. A, et al.  2003, ApJ, 592, 819

\bibitem[Bogdan et al. (2018)]{blk18} Bogdan, A. et al. 2018, ApJ, 869, 105

\bibitem[Bressan et al. (2006)]{bpb06} Bressan, A. Panuzzo, P., Buson, L., et al. 2006, ApJL, 639, L55

\bibitem[Brinchmann et al. (2004)]{bcw04} Brinchmann, J., Charlot, S., White, S. D. M., et al. 2004, MNRAS, 351, 1151

\bibitem[Bruzual \& Charlot (2003)]{bc03} Bruzual, G. \& Charlot, S. 2003, MNRAS, 344, 1000

\bibitem[Burns et al. (2008)]{bhg08} Burns, J. O., Hallman, E. J., Brennan, G., Motl, P. M., \& Norman, M. L. 2008, ApJ, 675, 1125

\bibitem[Cava et al. (2009)]{cb09} Cava et al. 2009, A\&A, 495, 707

\bibitem[Chabrier (2003)]{c03} Chabrier, G., 2003, PASP, 115, 763

\bibitem[Chang et al. (2015)]{cvc15} Chang, Y.-Y., van der Wel, A., da Cunha, E., \& Rix, H.-W. 2015, ApJS, 219, 8

\bibitem[Conselice, Gallagher \& Wyse (2001)]{cgw01} Conselice, C. J., Gallagher, J. S., III, \& Wyse, R. F. G. 2001, ApJ, 559, 791

\bibitem[da Cunha, Charlot, \& Elbaz (2008)]{dce08} da Cunha, E., Charlot, S., \& E \\lbaz, D. 2008, MNRAS, 339, 1595

\bibitem[Davari et al. (2017)]{dhm17} Davari, R. H., Ho, L. C., Mobasher, B., \& Canalizo, G. 2017, ApJ, 836, 75

\bibitem[de Jong (1996)]{dj96} de Jong, R. S. et al. 1996, A\&A 313, 377

\bibitem[Dekel \& Birnboim (2006)]{db06} Dekel, K. \& Birnboim, Y. 2006, MNRAS, 368, 2

\bibitem[Edge et al. (2003)]{ess03} Edge, A. C. et al. 2003, ApJ, 599, L69

\bibitem[Egami et al. (2006)]{emr06} Egami, E. et al. 2006, ApJ, 647, 922

\bibitem[Eisenstein et al. (2001)]{eag01} Eisenstein, D. J. et al. 2001, AJ, 122, 2267

\bibitem[Elbaz et al. (2007)]{edl07} Elbaz, D., Daddi, E., Le Borgne, D. et al. 2007, A\&A, 468, 33

\bibitem[Faisst et al. (2017)]{fcc17} Faisst, A. L. et al. 2017, ApJ, 839, 71

\bibitem[Falco et al. (1999)]{fk99} Falco et al. 1999, PASP, 111, 438

\bibitem[Fraser-McKelvie et al. (2016)]{fbp16} Fraser-McKelvie, A., Brown, M. J. I., Pimbblet, K. A., Dolley, T., Crossett, J. P., \& Bonne, N. J. 2016, MNRAS, 4612, L11

\bibitem[Governato et al. (2007)]{g07} Governato, F. et al. 2007, MNRAS, 374, 1479


\bibitem[Guzzo et al. (2009)]{gs09} Guzzo et al. 2009, A\&A, 499,  357

\bibitem[Hernandez-Toledo (2010)]{ht10} Hernandez-Toledo et al. 2010, AJ, 139, 2525

\bibitem[Hopkins et al. (2010)]{hbc10} Hopkins, P. F. et al. 2010, ApJ, 715, 202

\bibitem[Hopkins et al. (2009)]{hcy09} Hopkins, P. F., Cox, T. J., Younger, J. D., \& Hernquist, L. 2009, ApJ, 691, 1168

\bibitem[Hopkins et al. (2006)]{hhc06} Hopkins, P. F., Hernquist, L., Cox, T. J., Di Matteo, T., Robertson, B., \& Springel, V. 2006, ApJS, 163, 1

\bibitem[Lin et al. (2010)]{ls10} Lin, Y.-T., Shen, Y., Strauss, M. A., Richards, G. T., \& Lunnan, R., 2010, 

\bibitem[Lintott et al. (2011)]{lsb11} Lintott, C. J. et al. 2011, MNRAS, 410, 166

\bibitem[Lintott et al. (2008)]{lss08} Lintott, C. J., Schawinski, K., Slosar, A., et al. 2008, MNRAS, 389, 1179

\bibitem[Lotz et al. (2008)]{ljc08} Lotz, J. M.,  Jonsson, P., Cox, T. J., \& Primack, J. R. 2008, MNRAS, 391, 1137

\bibitem[Masters et al. (2010)]{mmr10} Masters, K. L. et al. 2010, MNRAS, 405, 783

\bibitem[Martig et al. (2009)]{mbt09} Martig, M., Bournaud, F., Teyssier, R., \& Dekel, A. 2009, ApJ, 707, 250

\bibitem[Melnick \& De Propris (2013)]{mp13} Melnick, J. \& De Propris, R. 2013, MNRAS, 431, 2034


\bibitem[Ogle et al. (2016)]{oln16} Ogle, P. M., Lanz, L., Nader, C., \& Helou, G. 2016, ApJ, 817, 109

\bibitem[Posti, L., Fraternali, F., \& Marasco, A. (2019)]{pfm19} Posti, L, Fraternali, F., \& Marasco, A., in press (arXiv:1812.05099)

\bibitem[Rines et al. (2001)]{rg01} Rines et al. 2001, ApJ, 561, 41

\bibitem[Rykoff et al. (2012)]{rk12} Rykoff et al. 2012,  ApJ, 746, 178

\bibitem[Rodriguez-Gomez et al. (2016)]{rg16} Rodriguez-Gomez, V. et al. 2016, MNRAS, 458, 2371 

\bibitem[Rowan-Robinson et al. (2005)]{rr05} Rowan-Robinson, M. et al. 2005, AJ, 129, 1183

\bibitem[Schawinski et al. (2014)]{sus14} Schawinski, K., Urry, C. M., Simmons, B. D., et al. 2014, MNRAS, 440, 889

\bibitem[Schlafley \& Finkbeiner (2011)]{sf11} Schlafly, E. F. \& Finkbeiner, D. P. 2011, ApJ, 737, 103

\bibitem[Sheth et al. (2008)]{see08} Sheth, K. et al. 2008, ApJ, 675, 1141

\bibitem[Simard et al. (2011)]{smp11}  Simard, L., Mendel, J. T., Patton, D. R., Ellison, S. L., \& McConnachie, A. W. 2011, ApJS, 196, 11



\bibitem[Skrutskie et al. (2006)]{s06} Skrutskie, M. F. et al. 2006, AJ, 131, 1163


\bibitem[Springel \& Hernquist (2005)]{sh05} Springel, V., \& Hernquist, L. 2005, ApJ, 622, L9 

\bibitem[Strauss et al. (2002)]{s02} Strauss, M. A., Weinberg, D. H., Lupton, R. H., et al. 2002, AJ, 124, 1810

\bibitem[Tremonti et al. (2004)]{t04} Tremonti, C. A. et al, 2004, ApJ, 613, 898

\bibitem[Veron-Cetty et al. (2004)]{vc04} Veron-Cetty et al. 2004, A\&A, 414, 487

\bibitem[Wright et al. (2010)]{w10} Wright, E. L. et al. 2010, AJ, 140, 1868

\bibitem[York et al. (2000)]{y00} York, D. G. et al. 2000, AJ, 120, 1579

\end{thebibliography}
\end{document}